\documentclass[conference]{IEEEtran}
\IEEEoverridecommandlockouts
\usepackage{cite}
\usepackage{algorithmic}
\usepackage{graphicx}
\usepackage{textcomp}
\usepackage{xcolor}

\usepackage{bm}
\usepackage{graphicx}
\usepackage[utf8]{inputenc}
\usepackage{mathtools}
\usepackage{amssymb}
\usepackage{xspace, amsthm, bbm, bm,theoremref}
\usepackage{paralist}
\usepackage{hyperref}

\usepackage{algorithmic}
\usepackage{multirow}
\usepackage{enumitem}
\usepackage{rotating}
\usepackage{array,makecell,multirow}
\usepackage{courier}  
\usepackage[caption=false]{subfig}

\newtheorem{theorem}{Theorem}

\newtheorem{definition}{Definition}

\usepackage{xspace}

\usepackage{algorithm}
\usepackage{multirow}
\usepackage{wrapfig,lipsum,booktabs}

\makeatletter
\def\ps@IEEEtitlepagestyle{%
  \def\@oddfoot{\mycopyrightnotice}%
  \def\@evenfoot{}%
}
\def\mycopyrightnotice{%
  {\footnotesize 978-1-6654-8045-1/22/$\$$31.00 ©2022 IEEE\hfill}
  \gdef\mycopyrightnotice{}
}

\def\BibTeX{{\rm B\kern-.05em{\sc i\kern-.025em b}\kern-.08em
    T\kern-.1667em\lower.7ex\hbox{E}\kern-.125emX}}

\title{AdvCat: Domain-Agnostic Robustness Assessment for
Cybersecurity-Critical Applications with Categorical Inputs
}

\author{\IEEEauthorblockN{
Helene Orsini\IEEEauthorrefmark{1}\IEEEauthorrefmark{2}\thanks{\IEEEauthorrefmark{1}The first four authors contributed equally.},
Hongyan Bao\IEEEauthorrefmark{1}\IEEEauthorrefmark{3},
Yujun Zhou\IEEEauthorrefmark{1}\IEEEauthorrefmark{3}, 
Xiangrui Xu\IEEEauthorrefmark{1}\IEEEauthorrefmark{4},\\
Yufei Han\IEEEauthorrefmark{2},
Longyang Yi\IEEEauthorrefmark{4},
Wei Wang\IEEEauthorrefmark{4}, 
Xin Gao\IEEEauthorrefmark{3},
Xiangliang Zhang\IEEEauthorrefmark{3}\IEEEauthorrefmark{5}\thanks{\IEEEauthorrefmark{5}Corresponding author.}
}
\IEEEauthorblockA{\IEEEauthorrefmark{2}\textit{INRIA}, France~
\IEEEauthorrefmark{3}\textit{King Abdullah University of Science and Technology}, Thuwal, SA}
\IEEEauthorblockA{\IEEEauthorrefmark{4}\textit{Beijing Jiaotong University}, Beijing, CN}
\IEEEauthorblockA{\IEEEauthorrefmark{5}\textit{University of Notre Dame}, Indiana, USA}
\IEEEauthorblockA{helene.orsini@irisa.fr, \{hongyan.bao,yujun.zhou\}@kaust.edu.sa, xiangrui.xu@bjtu.edu.cn,\\ yufei.han@inria.fr, \{longyang.yi,wangwei1\}@bjtu.edu.cn, xin.gao@kaust.edu.sa, xzhang33@nd.edu}
}

\begin{document}
\maketitle

\begin{abstract}
Machine Learning-as-a-Service systems (MLaaS) have been largely developed for cybersecurity-critical applications, such as detecting network intrusions and fake news campaigns. Despite effectiveness, their robustness against adversarial attacks is one of the key trust concerns for MLaaS deployment. We are thus motivated to  assess the adversarial robustness of the Machine Learning models residing at the core of these security-critical applications with categorical inputs. Previous research efforts on accessing model robustness {against manipulation of categorical inputs}  are specific to use cases and heavily depend on domain knowledge, {or require white-box access to the target ML model}. 
{Such limitations prevent the robustness assessment from being  as a domain-agnostic service provided to various real-world applications.} We propose a provably optimal yet computationally highly efficient adversarial robustness assessment protocol for a wide band of ML-driven cybersecurity-critical applications. We demonstrate the use of the domain-agnostic robustness assessment method with substantial experimental study on fake news detection and intrusion detection problems. 

\end{abstract}


\maketitle
\section{Introduction}\label{sec:intro}
Witnessed during recent decades, Machine Learning (ML)-based  techniques have powered increasingly more cybersecurity-critical services, such as intrusion detection and online misinformation investigation. The ML-based decision output in these applications directly impact human security analysts' judgment and strategic actions, like preventing underlying high-stake cyber intrusions and poisoning fake news 
\cite{wang2017liar} spreading in the early stage. 
However, a seemingly accurate ML-based detector/classifier may be highly vulnerable to adversarial input perturbations, which result in misdetection of security incidents. The lack of robustness against such input perturbations raises  trust concerns over the use of ML models in  cybersecurity-critical applications \cite{Pierazzi2020IEEESP}. Therefore, to provide trustable ML-as-a-Service  in these applications, it has become a must to assess the adversarial robustness of the core ML models, beyond achieving accurate detection/classification. \textit{The robustness assessment discloses to what extent the performance  of a core ML model can be affected by adversarial attacks at what expenses}.

There are three unaddressed key challenges in approaching the adversarial robustness assessment target in these cybersecurity-critical scenarios. 
\textbf{First}, the previous gradient-based approaches on understanding the robustness of deep learning models focus on continuous input \cite{Goodfellow2015}. 
They are ineffective and infeasible for the ML models with non-numerical categorical inputs. 
However, those categorical inputs pervasively exist in cybersecurity-critical applications \cite{shu2020fakenewsnet,deeplog}. For instance, intrusion detection identifies cyber attacks via a rich set of categorical behavioral signatures, such as malware execution traces, malicious network communication logs, and system event logs \cite{tiresias,deeplog}.
Misinformation detection is often based on learning the relationship of various grammatical categories such as word co-occurrences, subject-predicate-object triples, and writing styles \cite{shu2020fakenewsnet,wang2017liar}.
Unlike continuous measurements such as pixel intensities, categorical inputs have a limited number of options for the category values and have no intrinsic ordering to the categories. Conducting adversarial perturbations on categorical data are in nature an \textit{NP-hard knapsack problem} \cite{wang2020attackability}. As such, it is difficult to define a computationally efficient strategy to produce adversarial perturbations over categorical inputs.

\textbf{Second}, heuristic rules or trial-and-error processes may lead to feasible perturbations with additional assumptions narrowing the combinatorial perturbation space \cite{li2020bert-attack}. These methods nonetheless depend heavily on domain-specific knowledge and constraints defined by semantic integrity and rules. 
Such dependency limits the adaptive potential of the domain-specific assessment protocols to new domains. Moreover, domain-specific knowledge may not be always readily available. For example, the threat settings of cyber attacks vary drastically across different attack techniques and IT system architectures \cite{tiresias,deeplog}. Encoding domain-specific contexts of various intrusion incidents require expensive investigation overheads on a case-by-case basis. Besides, 
system threats may stay unknown to security analysts when an attack is delivered. It is impossible to define domain-specific rules for the zero-day attack events. The absence of \emph{a principled and domain-agnostic robustness assessment protocol} makes it difficult to provide an assessment-as-a-service pipeline to measure the adversarial vulnerability of different cybersecurity-critical applications.

\textbf{Third}, providing robustness assessment-as-a-service requires \emph{blackbox access} to the target ML model for safeguading the intellectual property. However, most existing attack methods require computing the gradient-based estimator to seek optimal adversarial data manipulations \cite{li2020bert-attack,wang2020attackability,QiSysML2018}, and thus require a \emph{white-box access} to the target ML model. Even though a few black-box adversarial attack methods \cite{Narodytska2017cvprw,Croce2019iccv,JDaaai2020,thorne2019adversarial} have been proposed, 
they either work on continuous data, or they are tailored to specific applications and can be barely extended across different domains. 

We are thus motivated to address these challenges and build \textbf{an out-of-the-box framework of Robustness Assessment-as-a-Service (RAaaS) on categorical inputs}. This framework presents an \textbf{automated and generally applicable robustness assessment pipeline} for any cybersecurity-critical domain with categorical inputs, even without significant domain-specific expertise. The produced diagnostic reports help benchmark the utility of the ML-assisted detection/classification under adversarial perturbations. 

{Our framework, named \textit{\textbf{AdvCat}}, employs three model-agnostic methods to solve the NP-hard knapsack problem of the robustness assessment task regarding categorical inputs, Forward  Stepwise  Greedy Search (\textit{FSGS}), Stochastic Greedy Search (\textit{SGS}), and  Upper-Confidence Bound Search (\textit{UCBS}) method. They conduct an iterative combinatorial search of feasible perturbations over categorical features, based on the classification output queried from the ML model. The diagnostic report generated by \textit{AdvCat} covers the performance change (e.g., drop of accuracy) caused by adversarial perturbation and the expense
of the robustness assessment process (e.g., the number of queries and runtime cost).} 
We highlight below the advantages making \textit{AdvCat} a unique and valuable asset to the community and summarize our contributions:
\begin{itemize}[leftmargin=*]
    \item \textit{AdvCat}, by design, is domain-agnostic and generally usable for any application with categorical inputs. However, if necessary, \textit{AdvCat} can be further customized to specific applications by integrating additional domain-dependent rules as plugins. Besides, \textit{AdvCat} provides 
    computationally efficient and provably accurate robustness assessment, which consolidates the use of \textit{AdvCat} across various ML models and application scenarios.
    \item \textit{AdvCat} only needs to query the decision output of the target ML model over a set of anonymized testing instances provided by the ML-driven application owner. This black-box setting safeguards data privacy and the intellectual property of ML models. 
    \item We demonstrate the workflow of \textit{AdvCat} on two cybersecurity-critical applications with categorical features: \textit{fake news detection} and \textit{intrusion detection}. Substantial experimental study unveils that 
    the state-of-the-art detection models in both cybersecurity practices are highly vulnerable to the discrete adversarial perturbations. The results raise serious concerns over the trustworthiness of the ML practices in detecting high-stake security events. 
\end{itemize}
It is worth noting that robustifying the vulnerable ML-driven cybersecurity applications is beyond the scope of our study. We therefore don't include robust training and other defense methods against adversarial perturbations in our empirical study. We discuss the related work discussion in Section.2, and introduce \textit{AdvCat} in Section.3. We establish utility analysis of \textit{AdvCat} in Section.4,  and conduct empirical study of two applications in Section.5. We conclude the whole paper in Section.6.
\section{Related work}\label{sec:related_work}

Tremendous efforts have been made to 
\emph{vulnerability measurement of a classifier under evasion attack} \cite{AFawzi2016nips,Matthias2017nips,cohen2019icml,weng2018icml,sinha2018iclr,shi2020robustness,Yin2018icml,TuNIPS19}.
Previous works focus on evaluating classifier robustness against $l_{p}$-norm perturbations on continuous data. They all assume adversarial samples locate within a smooth $l_{p}$-ball centered at an input instance, which doesn't hold for categorical data. 
Pioneering works of adversarial attacks on categorical data take advantage of domain-specific knowledge to facilitate the combinatorial search in categorical feature spaces. \cite{Yang2018GreedyAA,JFLindss2019,JDaaai2020,li2020bert-attack,zang2020word,wang2021adversarial,wang2020t3} focus on replacing words/sentences with manually chosen synonyms,  following syntactic guidelines or preserving entailment of texts to mislead text classifiers. 
\cite{Narodytska2017cvprw,Croce2019iccv} narrow down the search range within the image areas containing sensitive contents for image classification. Though the domain-specific constraints effectively improve the search efficiency of adversarial perturbations, the dependency on domain-specific knowledge prevents these evasion attack methods from being extended to new domains. Furthermore, the heuristic search strategies of these methods do not provide provably guarantees of the success of attack. 

Most adversarial attack methods on categorical data require white-box access to the target ML models, such as \textit{TextBugger} (the white-box setting) \cite{JFLindss2019}, \textit{BERT-Attack} \cite{li2020bert-attack}, \textit{SememePSO} \cite{zang2020word} and \textit{CompAttack} \cite{wang2021adversarial}. \cite{wang2020attackability,QiSysML2018} facilitate adversarial attacks on categorical data by computing gradients of embeddings of categorical features. \cite{bojchevskiICML19a,Bojchevski2019nips,BojchevskiICML2018} 
adopt edge-flipping and node attribute perturbation to poison graph neural networks (GNN). They appease the NP-hard problem by attacking surrogate models of GNN. These methods are tailored to specific GNN architectures. Though \cite{Narodytska2017cvprw,Croce2019iccv,JDaaai2020,JFLindss2019,thorne2019adversarial} adapt to the black-box attack setting, they depend heavily on domain knowledge. Notably, \cite{JDaaai2020} generates candidate words using specific word embedding models. \cite{thorne2019adversarial} requires human annotators to define entailment preserving rules to design adversarial texts. The lack of domain and model-agnostic adversarial perturbation methods becomes the bottleneck of building a generally applicable robustness assessment pipeline.

\cite{wang2020attackability} proposes a black-box and domain-agnostic attack method on categorical inputs, named \textit{FSGS}. It conducts iterative greedy search to select incrementally more categorical features to perturb. However, previous works \cite{wang2020attackability,QiSysML2018} only established the optimality of \textit{FSGS} when the target classifier has non-negative parameters, which is an unrealistic setting for ML practices.  Besides, the computational overheads of \textit{FSGS} increase drastically with higher feature dimensions. Our work answers the open problems. We unveil the rationality of using \textit{FSGS} in \textit{AdvCat} to evaluate the robustness of a general classifier. We further propose two alternative methods in \textit{AdvCat} to provide more economic yet accurate robustness assessment results. 
\begin{figure*}[t!]
    \centering
    \includegraphics[width=0.75\textwidth,height=3.5cm]{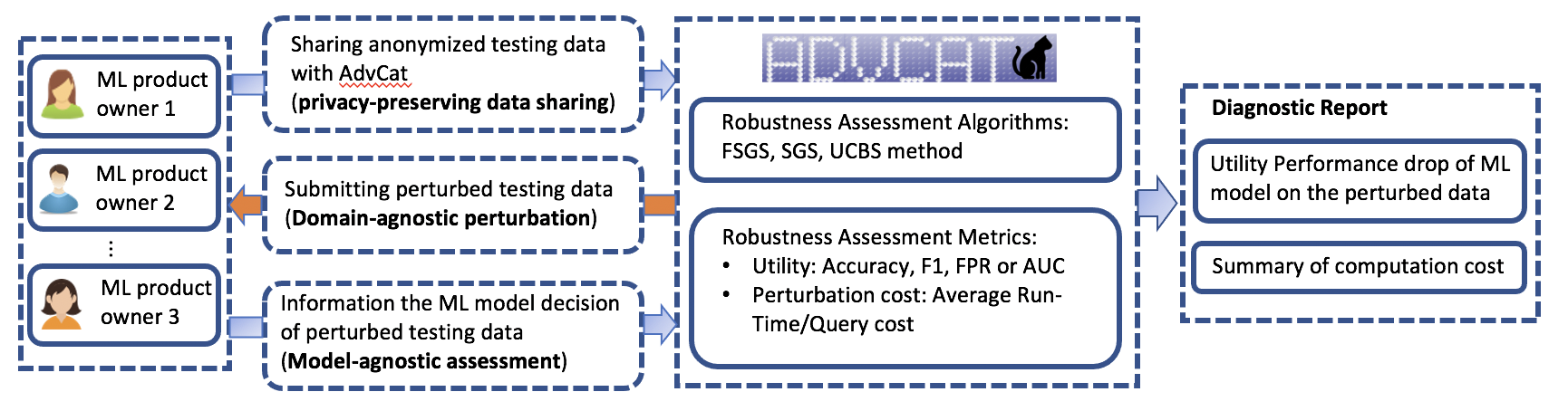}
    \vspace{-0.1cm}
    \caption{The workflow of \textit{AdvCat} for Robustness Assessment-as-a-Service }\vspace{-2mm}
    \label{fig:flowchat}
\end{figure*}

\section{AdvCat for Robustness Assessment }
\label{sec:system}

\subsection{Workflow of AdvCat}\label{sec:notation}
\textit{AdvCat} provides Robustness Assessment-as-a-Service to ML product owners, e.g., cybersecurity enterprises applying ML-driven intrusion or misinformation detectors. 
{We next introduce the workflow of \textit{AdvCat}, as illustrated in Figure.\ref{fig:flowchat}.} 

\noindent\textbf{Privacy-preserving data sharing.} An ML service provider shares with \textit{AdvCat} a set of testing instances, which will be used for empirically evaluating the adversarial robustness of the target ML model. The physical sense and concrete  values of each categorical feature are completely removed to prevent possible data privacy leak. The categorical values are replaced by integer indexes.
 
\noindent \textbf{Model-agnostic assessment.} Instead of sharing the ML model, the product owner provides the query access, e.g. query APIs to \textit{AdvCat}. It allows \textit{AdvCat} to get the predicted class label and the classification confidence of a submitted testing instance. 
We leave the label-only scenarios for future study. Besides, we don't assume the availability of any surrogate model to the target ML model as in \cite{wang2021adversarial}. Building surrogate models requires extra efforts to collect shadow training data and prior knowledge of the target model architecture, which violates the model-agnostic setting of \textit{AdvCat}. 

\noindent\textbf{Domain-agnostic perturbation.} The core of \textit{AdvCat} is to construct perturbed instances and test if they can bypass the detector. \textit{AdvCat} constructs adversarial perturbations only based on the queried decision output from the ML model via the API access, without using any domain-specific knowledge. We introduce the domain and model-agnostic assessment methods in Section 3.3. Based on the classification results of perturbed instances, \textit{AdvCat} generates diagnostic reports profiling the adversarial robustness level of the core ML model of the application. 
\subsection{Notations and Problem Definition}
Let $f$ denote the ML classifier of a cybersecurity-critical application, which predicts a label $y_k$ ($k=1,...,{K}$, $K\geq{2}$) for an instance $\bm{x} = \{x_1,x_2,...,x_n\}$ of $n$ categorical attributes. Each of the categorical attributes $x_i$ takes any of ${m}$ (${m}\geq{1}$) categorical values. In real-world cybersecurity applications, $x_i$ can be one of the behavioural signatures associated with network attacks or keywords of fake news texts.   
In practices, we cast each categorical value of $x_i$ to a $D$-dimensional pre-trained embedding vector, e.g., $\bm{e}^j \in{\mathbb{R}^{D}},\, j=1,2,...,m$. 
To represent an instance $\bm{x}$ with the embedding vectors of its category values, we define binary variables $\bm{b} = \{b^{j}_{i}\}$, $i$=$1,2,...,n$, $j$=$1,2,...,m$, where  $b^{j}_{i} = 1$ when the $j$-th feature value is present for $x_i$ and $b^{j}_{i} = 0$ otherwise. An instance $\bm{x}$ can then be represented by an  $\mathbb{R}^{n*m*D}$ tensor with ${x}_{\{i,j,:\}} = b^{j}_i{\bm{e}^{j}_i}$. 

We use $\bm{\hat{b}} = \{{\hat{b}}^{j}_{i}\}$ to indicate the adversarial modifications introduced into $\bm{b}$. For unperturbed $x^{j}_{i}$, ${\hat{b}}^{j}_{i}= {{b}}^{j}_{i}$. Otherwise, ${\hat{b}}^{j}_{i}\neq{{b}}^{j}_{i}$. Depending on the type of attacks to implement (\emph{insertion}, \emph{deletion} or \emph{substitution}), $\hat{b}^{j}_{i}$ can have different values.
Although $f$ is unknown in our setting, \textit{AdvCat} can query the decision probabilities $f_{y_k}(\bm{x},\bm{\hat{b}})$ of instance $\bm{x}$ with perturbation $\bm{\hat{b}}$. 
For a target ML model $f$, the problem of evaluating the adversarial robustness on a given  input $\bm{x}$ can be formulated by \textit{set function maximization} in Eq.\ref{eq:optimal_attack}. 
\begin{equation}\label{eq:optimal_attack}
\small
\psi(l) = \underset{\substack{\hat{{b}}_{i=1...n},\\ l= \textit{diff}({b},\hat{{b}})}}{\max} \quad \underset{\hat{b}_i^{j=1...m} }{\max} (m_f)  \qquad   \text{s.~t.} \quad m_f \geq{0},\,\;\;|l|\leq \varepsilon
\end{equation}
where $m_f=\underset{k\in\{1,...,{K}\}/\{k^*\}}{\max}\{f_{y_k}(\bm{x},\hat{\bm{b}})\} - f_{y_{k^*}}({\bm{x}},\hat{\bm{b}})$ defines the decision confidence gap over an input instance $\bm{x}$ between  the wrong and correct classification output. $k^*$ is the ground truth label. And $m_{f}\in[-1,0)$ and $m_{f}\geq{0}$ correspond to a correct and wrong classification, respectively.  Note that we focus on the resilience against \textit{non-targeted evasion attack} in this work. The threat of targeted attack  considers additional rank constraints on the classifier's output over different classes, and will be investigated in future. 

We aim at seeking a \textbf{minimal set of categorical feature perturbations $l= \textit{diff}(\bm{b},\hat{\bm{b}})$}, with which the \textbf{maximum possible difference} between the decision confidence of any wrong class label $k$ and the correct label ${k^*}$ is larger than 0. 
The upper limit of the size of the perturbation set $l$, noted as $\varepsilon$, is used as the budget of adversarial perturbation and a tunable parameter of the assessment service. Higher/lower $\varepsilon$ indicates stronger/weaker strength of input perturbation. \textit{AdvCat} adapts $\varepsilon$ to customize the perturbation strength in different applications. Given a target ML model, solving Eq.\ref{eq:optimal_attack} to induce $m_{f}\geq{0}$ provokes an alert of flipped decision output on the disturbed input. More/less flipped decisions over the given testing set indicates lower/higher adversarial robustness of the target ML-driven application. 
\subsection{Benchmark Algorithms in \textit{AdvCat}}
To solve Eq.\ref{eq:optimal_attack}, \textit{AdvCat} proposes to use three optional algorithms (\textit{FSGS}, \textit{SGS} and \textit{UCBS}) 
to address the NP-hard problem of finding a minimal set of  $l= \textit{diff}(\bm{b},\hat{\bm{b}})$. Given an input $\bm{x}$, they conduct iterative combinatorial search of feasible perturbations $\hat{\bm{b}}$ over categorical features, based on the classification output queried from the ML model. They stop the search at iteration $t$ when $m_{f}(l_{t})\geq{0}$ (achieving a successful attack with  $l_t$) or the number of modifications $|l_t| > \varepsilon$. To control the attack processing time in practice, we also set a run-time limit (denoted as $T_{L}$) to stop the search if no feasible perturbations can be found to make $m_{f} \geq{0}$ for input $\bm{x}$.      

\noindent\textbf{Forward Stepwise Greedy Search (\textit{FSGS}).}
Given a categorical input $\bm{x}$, let $S$ denote the support set of perturbed categorical features, which is initially set as an empty set. In each iteration of \textit{FSGS}, we expand $S$ by adding one candidate feature of $\bm{x}$ that can bring the maximum increase of $m_{f}$ by perturbing the combination of this feature and the features already in $S$. More formally, Algorithm \ref{alg:FSGS}   gives the pseudo codes of \textit{FSGS}. In each iteration $t$, \textit{FSGS} first runs the \emph{inner level optimization} (Line 5). For each candidate feature $a_{i}$, this is to locate the union of a subset of $S_{t-1}$ and $a_{i}$ that can induce the maximum increase of $m_{f}$. Then in the \emph{outer level optimization}, \textit{FSGS} selects the  candidate feature $\tilde{a}$ and adds it to $S_{t}$, if the perturbation over $S_{t}\cup{\tilde{a}}$ can achieve the largest marginal gain of $m_{f}(S_{t}\cup{\tilde{a}})$ over $m_{f}(S_{t-1})$ (Line 8).

\begin{algorithm}[t]
 \caption{FSGS algorithm for robustness assessment}
 \label{alg:FSGS}
 \small
 \begin{algorithmic}[1]
  \REQUIRE 
   The input $\bm{x}$, the modifiable features $H$=$\{a_{i}\}_{i=1...n}$, the attack budget $\varepsilon$, the run-time limit $T_L$
  \ENSURE  The maximum support set $S_{t}$ pushing $m_{f}(l_{t})$ to violate the tolerance constraint in Eq.~\ref{eq:optimal_attack}
  
  \STATE $S_{0}\leftarrow{\emptyset}$,~Run-time $T_R = 0$
  \FOR{$t=0,1,2,\ldots$}
    \STATE $T\leftarrow{\textit{zero-valued vector} \in{\mathbb{R}^{n}}}$\\
    \FOR{each\,\,$a_{i}\in{H/S_{t-1}}$}
        {
          \STATE $T(a_{i})$=$\underset{s\subset{S_{t-1}}, |s|\leq \varepsilon}{\arg\max} m_{f}(s\cup{a_{i}})$
          $(\textbf{\textit{Inner\,\,Level\,\,Optimization}})$\\ 
      }
    \ENDFOR
    \STATE $\tilde{a} = \underset{a_{i}}{\arg\max} \;\;m_f(T(a_{i})\cup {a_i})\;(\textbf{\textit{Outer\,\,Level\,\,Optimization}})$\\
    \STATE $S_{t}{\leftarrow}S_{t-1}\cup{\{\tilde{a}\}}$\\
    \STATE $l_t=T(\tilde{a})\cup{\tilde{a}}$
     
    \STATE \textbf{if} {$m_{f}(l_{t})\geq{\Gamma}$ ~or~Run-time $T_R > T_L$ }  \textbf{then break}

  \ENDFOR
 \end{algorithmic}
\end{algorithm}

\noindent\textbf{Stochastic Greedy Search (\textit{SGS}).}
This is a stochastic variant of the greedy search solution \cite{Chen2017WeaklySM}. It has a lower time complexity than \textit{FSGS}. In each iteration of \textit{SGS}, instead of traversing all the candidate features not in $S$, we randomly choose $r$ of the candidate features. Then we expand the set $S$ by adding one of the $r$ candidate features to reach the largest marginal gain, as \textit{FSGS} does. The details of \textit{SGS} are presented in Algorithm \ref{alg:SGS}.

\begin{algorithm}[t]
 \caption{SGS algorithm for robustness assessment}
 \small
 \label{alg:SGS}
 \begin{algorithmic}[1]
  \REQUIRE 
  The input $\bm{x}$, the modifiable features $H$=$\{a_{i}\}_{i=1...n}$, the number of  features to randomly select $r$, the attack budget $\varepsilon$, the run-time limit $T_L$
  \ENSURE  The maximum support set $S_{t}$ pushing $m_{f}(l_{t})$ to violate the tolerance constraint in Eq.~\ref{eq:optimal_attack}
  
  \STATE $S_{0}\leftarrow{\emptyset}$,~Run-time $T_R = 0$
  \FOR{$t=0,1,2,\ldots$}
    \STATE $T\leftarrow{\textit{zero-valued vector} \in{\mathbb{R}^{n}}}$\\
    \STATE $R\leftarrow{\text{randomly choose $r$ features from}\;\; H/S_{t-1}}$
    \FOR{each\,\,$a_{i}\in R$}
        {
          \STATE $T(a_{i})$=$\underset{s\subset{S_{t-1}}, |s|\leq \varepsilon}{\arg\max} m_{f}(s\cup{a_{i}})$
          $(\textbf{\textit{Inner\,\,Level\,\,Optimization}})$\\ 
      }
    \ENDFOR
    \STATE $\tilde{a} = \underset{a_{i}}{\arg\max} \;\;m_f(T(a_{i})\cup {a_i})\;(\textbf{\textit{Outer\,\,Level\,\,Optimization}})$\\
    \STATE $S_{t}{\leftarrow}S_{t-1}\cup{\{\tilde{a}\}}$\\
    \STATE $l_t=T(\tilde{a})\cup{\tilde{a}}$
     
    \STATE \textbf{if} {$m_{f}(l_{t})\geq{\Gamma}$ ~or~ Run-time $T_R \geq T_L$ }  \textbf{then break}

  \ENDFOR
 \end{algorithmic}
\end{algorithm}

\noindent\textbf{Upper-Confidence Bound Search (\textit{UCBS}).}  Different from \textit{FSGS} and \textit{SGS}, this method adopts the criterion of Upper-Confidence Bound (UCB) {\cite{auer2002finite}} to explore the combinatorial space of adversarial perturbations. It is a popular Multi-Armed Bandit (MAB) technique, which explicitly balances the exploration and exploitation of the combinatorial search {\cite{auer2002finite}}. 
To employ \textit{UCBS} for robustness assessment, we consider one \emph{categorical feature to  perturb} in  $\bm{x}$ as  \emph{one arm to pull} in MAB. Unlike \textit{FSGS} and \textit{SGS}, \textit{UCBS} does not need to evaluate the effect of  perturbations over all possible combinations of the selected features in each iteration. 
As  given in  Algorithm \ref{alg:UCB}, the first step is to obtain the initial reward information after modifying each feature $a_i$ once (line 2-3 in Algorithm \ref{alg:UCB}). 
The reward $G_{a_i,t}$ is determined by the classifier's prediction on the implemented modification at iteration $t$, including $a_i$ as well:
\begin{equation}
\small
    G_{a_i,t} = \underset{k\in\{1,...,{K}\}/\{k^*\}}{\max}\{f_{y_k}(\bm{x},\hat{\bm{b}}_{a_i,t})\} .
\end{equation}
The reward $G_{a_i,t}$ has a value in $[0,1]$. If the feature $a_i$ is not selected in iteration $t$, the reward $G_{a_i,t}$ will be 0.
Over all previous  $t-1$ iterations,  the experimentally expected reward of perturbing  feature $a_i$ is:
\begin{equation}
\small
    \hat{\mu}_{a_i,t-1} = \frac{1}{t-1} \sum_{\tau=1}^{t-1} G_{a_i,\tau}.
\end{equation}
\vspace{-1mm}
Then $\hat{\mu}_{a_i,t-1}$ is used to build the upper confidence bound function,
\begin{equation}\label{eq:UCBv}
\small
B_{a_i,T_i,t-1} = \hat{\mu}_{a_i,t-1} + \sqrt{\frac{ \alpha *\log{t} }{2T_{a_i,t-1}} },
\vspace{-1mm}
\end{equation}
\vspace{-1mm}
$\alpha$ is the parameter to adjust the exploration depth and exploitation level. $T_{a_i,t-1}$ is the accumulated number that the feature $a_i$ been selected during the whole $t-1$ iterations. 

The second step is the iteration searching process. At each iteration, it will select one feature $\tilde{a}$ with the maximum value of $B_{a_i,T_i,t-1}$ to update $S_t$ (line 6-7 in Algorithm \ref{alg:UCB}). After several iterations of \textit{UCBS}, the combination of the selected features from the modifiable features could cause miss-classification/miss-detection output of the target ML model. 

\begin{algorithm}[t]
\small
\caption{The UCBS algorithm for robustness assessment }\label{alg:UCB} 
\begin{algorithmic}[1]
\REQUIRE
The input $\bm{x}$, the modifiable features $H$=$\{a_{i}\}_{i=1...n}$, the attack budget $\varepsilon$,   the run-time  limit $T_L$\,\, 
\ENSURE
the selected features $S_{t}$ to perturb;\\
\STATE $S_{0}\leftarrow{\emptyset}$, $t = 0$, Run-time $T_R = 0$\\
\FOR{$i_0\in{H}$}{
           \STATE $G_{a_i,0} {\leftarrow} {a_{i,0}} $ $~~~~~~~~~~~~~~~~~~~$ modify each feature once\\
          }
          \ENDFOR \\
\WHILE{$m_{f}(S_{t})\geq{0}$ or $|S_t| \le \varepsilon $ and $T_R$ $\leq$ ${T_L}$}{
          {
                 \STATE ${\tilde{a}} = {\underset{a_i\in{H}}{\arg\max}{\; \hat{\mu}_{a_i,t-1} + \sqrt{\frac{ \alpha *\log{t} }{2T_{a_i,t-1}} }}} $ \\
                 \STATE ${S_{t}~~{\leftarrow}~~ S_{t-1}~{\cup}~{\tilde{a}}}$\\
                 \STATE Update reward ${G_{a_i,t}}$,~ $m_{f}(S_{t})$,~ Run-time $T_R$\\
                 \STATE t = t + 1 \\
              }
      }
      \ENDWHILE

\end{algorithmic}
\end{algorithm}

\noindent\textbf{Assessment Algorithm Discussion.} Previous studies \cite{Croce2019iccv,QiSysML2018,wang2020attackability} lack the optimality guarantee of using \textit{FSGS} to evaluate the robustness of a general classifier. It prevents the use of \textit{FSGS} for domain-agnostic robustness evaluation. We address the limit by establishing the provably optimality of \textit{FSGS} for general ML models in Section \ref{sec:assess_algo_analysis}. 

To appease the concern over the intense computational cost of \textit{FSGS} over high-dimensional categorical feature space, we propose \textit{SGS} and \textit{UCBS} as two computationally economic alternatives for robustness assessment. These two methods are adopted for adversarial robustness assessment for the first time. As reported in Section  \ref{sec:assess_algo_analysis}, \textit{SGS} and \textit{UCBS} are significantly more efficient than \textit{FSGS} in solving the NP-hard knapsack problem. In addition, the quality analysis in Section  \ref{sec:assess_algo_analysis} shows that they can also achieve provably accurate assessment results, as in \textit{FSGS}.

As shown in Algorithms 1-3, the three methods can take additional domain-specific rules defining modifiable categorical features as optional inputs. In this way, \textit{AdvCat} can take domain-specific knowledge as plugin to be better customized to concrete applications.

\subsection{Robustness Metrics}
Given an ML-driven application, \textit{AdvCat} offers options to use \textit{FSGS}, \textit{SGS} or \textit{UCBS} method alternatively to search for feasible adversarial perturbations over each testing instance. \textit{AdvCat} then measures the security event detection performances before and after injecting the perturbations, in terms of the popular metrics including \textit{Accuracy} (\textit{Acc}), \textit{F1 Score} (\textit{F1}), \textit{AUC score} (\textit{AUC}) or 
\textit{False Positive Rates} (\textit{FPR}). The smaller/greater decrease of Acc, F1 and AUC after performing  perturbations indicates that the ML-driven application is more/less robust against the perturbations. In contrast,   a larger \textit{FPR} denotes less resilience to the  perturbations and vice versa. 

To illustrate the variance of these accuracy metrics caused by the attack, \textit{AdvCat} reports \textit{the percentage decrease / increase of the detection metrics} after the attack is enforced, compared to the adversary-free metrics. They include \textit{Difference of Acc} (\textit{DAcc}$\downarrow$), \textit{Difference of F1} (\textit{DF1}$\downarrow$), \textit{Difference of FPR} (\textit{DFPR} $\uparrow$), or \textit{Difference of AUC} (\textit{DAUC}$\downarrow$) 
respectively. $\downarrow$ / $\uparrow$
denotes the corresponding 
drop / rise 
after perturbation in \textit{AdvCat} compared with the unperturbated  performance. For intrusion detection, we consider additionally \textit{Detection Rate} (\textit{DR}) of abnormal system sessions and report \textit{Difference of DR} (\textit{DDR}$\downarrow$). 

\textit{AdvCat} also reports the expenses of robustness assessment over the testing instances. The expenses are measured by Averaged Query Number 
and {Averaged Running Time} (\textit{Runtime}), which are  
the averaged number of queries that \textit{AdvCat} sends to the ML API and the averaged runtime of perturbing {each testing instance}. Larger/smaller these meansured expenses
represent higher/lower overheads of robustness assessment. 
\section{Utility Analysis of \textit{AdvCat}}
\subsection{Query Complexity Analysis}
 In the process of robustness assessment, \textit{AdvCat} iterativelly queries the ML model to explore feasible categorical data manipulation. The query cost hence constitutes the majority of the runtime cost of \textit{AdvCat}. The more/less required queries indicate the higher/lower computational complexity of \textit{AdvCat} in practices. For one instance $\bm{x}$ containing $n$ categorical features and each feature having one of  $m$ categorical values, Table \ref{tab:complexity} reports the query complexity of \textit{FSGS}, \textit{SGS} and \textit{UCBS} running $T$ iterations.
The comparison shows that the greedy based attack methods  
(\textit{FSGS}, \textit{SGS}) have significantly higher query complexity than \textit{UCBS} method when the running iterations of $T$ and the number of features $n$ become larger. 
\begin{table}
\caption{Query Complexity of \textit{FSGS}, \textit{SGS} and \textit{UCBS}} 
\vspace{-2mm}
\hspace{-1pt}
\label{tab:complexity}
\small
\centering
\begin{tabular}{c|c}
\toprule
\textbf{Assessment method}        & \textbf{Query Complexity}\\ 
\midrule
FSGS          & $\sum_{t = 0}^T((n-t)*m*2^t)$      \\ 
SGS         & $\sum_{t = 0}^T(r*m*2^t)$       \\ 
UCBS    & $n*m + T$     \\ 
\bottomrule
\end{tabular} 
\vspace{-3mm}
\end{table} 

\subsection{Assessment Quality Analysis}\label{sec:assess_algo_analysis}
To provide consistently accurate robustness evaluation across different cybersecurity applications and different ML model architectures, it is   necessary to analyze the solution quality of \textit{AdvCat}  by \textit{FSGS}, \textit{SGS} and \textit{UCBS} in solving the  problem  in Eq.\ref{eq:optimal_attack}. We next establish the analysis that 
\textit{AdvCat} can provide provably accurate assessment results, regardless of the applications and the choice of ML models .

\begin{definition}\label{def:boundness}
\textbf{Smoothness of $f$.} Let $\Omega=({p},{q})$, ${p},{q}\in{\mathbb{R}^{n}}$ and the classifier $f$: $\mathbb{R}^{n}{\to}\mathbb{R}$ be a Lipschitz-continuous and differentiable function. $f$ is  $(m_{\Omega},M_{\Omega})$-\textit{smooth} on $\Omega$, if for any $({p},{q})\in\Omega$, $m_{\Omega}\in{\mathbb{R}}$ and $M_{\Omega} \in {\mathbb{R}^{+}}$,  $\epsilon = f({q})-f({p}) - \langle\nabla{f}({p}),{q}-{p}\rangle$ satisfies:
\begin{equation}\label{eq:smoothness}
\small
    \frac{m_{\Omega}}{2}\|{q}-{p}\|^2_{2} \leq |\epsilon| \leq  \frac{M_{\Omega}}{2}\|{q}-{p}\|^2_{2}.
\end{equation}
\end{definition}
According to \cite{jordan2021exactly}, most deep neural networks are practically Lipschitiz continuous. These popularly employed classifiers with finite Lipshitz constants meet the smoothness condition. $M_{\Omega_{\zeta}}$ and $m_{\Omega_{\zeta}}$ can be computed as local Lipshitz constant and the local strong convexity constant of $f$ respectively. Therefore, the smoothness condition allows much broader and more realistic choices of the classifier, compared to the non-negativity constraint proposed in \cite{QiSysML2018,wang2020attackability}. Given a classifier satisfying Eq.\ref{eq:smoothness}, Eq.\ref{eq:optimal_attack} defines in nature a weakly submodular optimization problem. The definition of weak submodularity can be found in \cite{elenberg2018restricted}.

\begin{theorem}\label{theorem:submod_max}
\textbf{Weak submodularity of the robustness assessment problem.} Let $\Omega_{\zeta} = \{(\hat{\bm{b}},\hat{\bm{b}'}):|\text{diff}\,(\bm{b},\hat{\bm{b}})|\leq{\zeta},|\text{diff}\,(\bm{b},\hat{\bm{b}}')|\leq{\zeta},|\text{diff}\,(\hat{\bm{b}},\hat{\bm{b}}')|\leq{\zeta},\,\zeta\geq{1}\}$, where $\hat{\bm{b}}$ and $\hat{\bm{b}}'$ denote two sets of  modifications over $\hat{\bm{b}}$. If the classifier $f_{y_k}(\bm{x},\bm{b})$  follows the regularity condition given by $(m_{k,\Omega_{\zeta}},M_{k,\Omega_{\zeta}})$-\textit{smoothness} constraint on $\Omega_{\zeta}$, the robustness assessment problem defined in Eq.\ref{eq:optimal_attack} can be formulated respectively as \textbf{monotone $\gamma_{\zeta}$-weakly submodular maximization}.

Let {\footnotesize $\epsilon_{k} = f_{y_k}(\bm{x},\hat{\bm{b}}') - f_{y_k}(\bm{x},\hat{\bm{b}}) - \langle\nabla{f}_{y_k}(\bm{x},\hat{\bm{b}}),\hat{\bm{b}}'-\hat{\bm{b}}\rangle$},  and {\footnotesize $\nabla{f}_{y_k}(\bm{x},\bm{b})_{\nu}$} denote the elements of {\footnotesize $\nabla{f}_{y_k}(\bm{x},\bm{b})$} corresponding to the difference between  $\hat{\bm{b}}$ and $\hat{\bm{b}}'$, where {\footnotesize $\nu = \textit{diff}(\hat{\bm{b}},\hat{\bm{b}}')$. }
The submodularity ratio $\gamma_{\zeta}$ for the robustness assessment problem of Eq.\ref{eq:optimal_attack} is bounded as:
\begin{equation}\label{eq:wsub_max}
\small
    \gamma_{\zeta} =  \underset{k=1,...,K}{\min}\,\{\gamma_{k,\zeta}\}  
\end{equation}
where {\footnotesize $\gamma_{k<K,\zeta}
    \geq\frac{\|\nabla{f}_{y_k}(\bm{x},\bm{b})_{\nu}\|_{2}+m_{k,\Omega_{1}}|\zeta|/2}{\|\nabla{f}_{y_k}(\bm{x},\bm{b})_{\nu}\|_{2}+M_{k,\Omega_{\zeta}}|\zeta|/2}$ } for {\footnotesize $\epsilon_{k}\geq{0}$,  \\
    $\gamma_{k<K,\zeta}
    \geq\frac{2m_{k,\Omega_{\zeta}}}{\|\nabla{f}_{y_k}(\bm{x},\bm{b})_{\nu}\|^2_{2}}(\|\nabla{f}_{y_k}(\bm{x},\bm{b})_{\nu}\|_{2}-M_{k,\Omega_{1}}|\zeta|/2)$ } for {\footnotesize $\epsilon_{k}<{0}$,  \\
    $\gamma_{k^*,\zeta} \geq \frac{2m_{K,\Omega_{\zeta}}}{\|\nabla{f}_{y_{k^*}}(\bm{x},\bm{b})_{\nu}\|^2_{2}}(\|\nabla{f}_{y_{k^*}}(\bm{x},\bm{b})_{\nu}\|_{2}-M_{K,\Omega_{1}}|\zeta|/2)$} for {\footnotesize $\epsilon_{k^*}\geq{0}$,}  and  \\ {\footnotesize
    $\gamma_{k^*,\zeta}
    \geq\frac{\|\nabla{f}_{y_{k^*}}(\bm{x},\bm{b})_{\nu}\|_{2}+m_{K,\Omega_{1}}|\zeta|/2}{\|\nabla{f}_{y_{k^*}}(\bm{x},\bm{b})_{\nu}\|_{2}+M_{K,\Omega_{\zeta}}|\zeta|/2}$ } for {\footnotesize $\epsilon_{k^*}<{0}$.}  
\end{theorem}

\begin{theorem}\label{theorem:fsgs_bound}
\textbf{Provably solution quality of \textit{FSGS} and \textit{SGS}.} For a $(m_{\Omega_{\zeta}},M_{\Omega_{\zeta}})$-\textit{smooth} classifier $f$, the quality of the solution $l$ to Eq.\ref{eq:optimal_attack} using \textit{FSGS} and \textit{SGS}  can be bounded respectively as 
\begin{equation}\label{eq:fsgs_bound}
\begin{split}
\small
    &|m^{\textit{FSGS}}_{f}(l)| \leq  e^{-{\gamma}_{\zeta}}+(1-e^{-{\gamma}_{\zeta}})|m_{f}(l^{\textit{OPT}})|\,\;\;\;\;\,\,\\
    &|m^{\textit{SGS}}_{f}(l)| \leq  1- (1+1/\gamma_{\zeta})^{-2} + (1+1/\gamma_{\zeta})^{-2}|m_{f}(l^{\textit{OPT}})|\,\;\;\;\;\,\,\\
\end{split}
\end{equation}
where $\gamma_{\zeta}$ is the submodularity ratio in Eq.~\ref{eq:wsub_max}. $l^{\textit{OPT}}$ is the underlying true optimal solutions to the robustness assessment problem of Eq.\ref{eq:optimal_attack}. 
\end{theorem}

\noindent{\textbf{Solution quality of FSGS and SGS}.} The theoretical analysis established in Theorem~\ref{theorem:submod_max} and \ref{theorem:fsgs_bound} shows the rationality of using \textit{FSGS} and \textit{SGS} in \textit{AdvCat} from two perspectives. \textbf{First}, if the target  model $f$ is $(m_{\Omega_{\zeta}},M_{\Omega_{\zeta}})$-\textit{smooth}, these two methods can provide good approximation quality to the underlying optimal result of the robustness assessment problem, regardless of the architecture of $f$, {as given in Eq.\ref{eq:fsgs_bound}}. 
\textbf{Second}, the solution quality guarantee of \textit{SGS} is less tight compared to that of \textit{FSGS} with the same submodularity ratio. It indicates that the accuracy of the \textit{SGS}-based robustness asseessment may practically present relatively larger variation than \textit{FSGS} across different applications. Nevertheless, in contrast to \textit{FSGS}, \textit{SGS} has a significantly reduced query cost. This makes \textit{SGS} more feasible for   scenarios where the query budget to the target ML-based application is limited. These two choices enable \textit{AdvCat} to make a trade-off between the theoretical optimality and the empirical feasibility of the robustness assessment task.  

\noindent{\textbf{Regret bound of UCBS}.} Unlike \textit{FSGS} and \textit{SGS}, the solution quality of \textit{UCBS}  is not built upon the weak submodularity nature of the robustness assessment problem. We establish the regret bound for \textit{UCBS} to analyse the quality of its solution to the robustness assessment problem in Eq.\ref{eq:optimal_attack}. The UCB regret generally indicates the difference between the selected condition and the ideal optimized condition. Concretely in our study, the regret $\bigtriangleup_{a_i}$ of \textit{UCBS} is defined as the gap between the adversarial perturbation effects using the selected features and the underlying optimal perturbation effects,
\begin{align}
\bigtriangleup_{a_i} & =\mu^* - \mu_{a_i}.
\end{align}
where $\mu^* = \max_{1\leq i \leq n}{\mu_{a_i}}$ is the largest expected reward.  
The expected regret $E[{R}_T]$ bound of \textit{UCBS} under $T$ iterations is \cite{auer2002finite}:
\begin{align}
E[{R}_T] \leq {\sum_{\bigtriangleup_{a_i} > 0} (\frac{2\alpha}{\bigtriangleup_{a_i}}\log T + \frac{\alpha}{\alpha -2}{\bigtriangleup_{a_i}}}).
\end{align}
The smaller the regret becomes, the more accurate the \textit{UCBS} robustness assessment result is.
The expected regret bound increases slowly as running more iterations $T$. The \textit{UCBS} can guarantee to search out the high quality feature chain.

\section{{AdvCat} on Security-critical Applications}


\subsection{Cybersecurity-critical Applications}

\noindent\textbf{Fake news detection.} 
In general, fake news detection models are built on a \emph{news embedding} + \emph{classification} pipeline.  Word2Vec \cite{mikolov2013efficient}, Glove \cite{pennington2014glove} and Bert \cite{devlin2018bert} are  most widely used for \emph{news embedding}, and CNN, LSTM, and MLP are employed for \emph{classification}. 
Our study involves the state-of-the-art detection models, including \textbf{CNN with Glove} \cite{wang2017liar}, \textbf{CNN with Bert}, \textbf{MLP with Word2Vec} \cite{dou2021user}, \textbf{MLP with Glove}, \textbf{MLP with Bert} \cite{dou2021user, wang2021adversarial, maheshwary2021generating}, \textbf{LSTM with Word2Vec}, \textbf{LSTM with Glove} \cite{wang2017liar, iyyer2018adversarial}, and \textbf{LSTM with Bert}.  
CNN and LSTM classifiers take input of words/sentences  embedding. 
MLP takes the embedding for the whole text as input. 
Few studies consider combinatorial attack of these adversarial words or sentences. 
In this work, we systematically study the combinatorial attack in fake news detection.

\noindent\textbf{Intrusion detection.}
Deep Learning-based intrusion detection methods \cite{deeplog,deepcase} aim to automatically flag suspicious attack events and category different attack techniques using network traffic logs. 
A vulnerable intrusion detection model to the adversarial perturbations can cause miss-detection of cyber attacks and expose the key infrastructure to the attackers. 
We involve a state-of-the-art Deep Neural Network-based solution to detecting abnormal events over security event logs, namely \textit{DeepLog} \cite{deeplog} in our benchmark study. \textbf{{DeepLog}} models the sequence of security event logs of normal system sessions using a Long Short-Term Memory (LSTM). 
Specifically, given the first $M$ log entries $\{e_{t-M},...,e_{t-2},e_{t-1}\}$ as input, \textit{DeepLog} leverages LSTM to predict the successive log $e_{t}$. By treating the integer index of the log-to-predict $e_{t}$ as a class label, \textit{DeepLog} conducts log prediction as multi-class classification. It adopts the top-$K$ classification scheme: it checks if the target log is one of the top $K$ predictions (the $K$ predicted log with the highest classification confidence). The log-prediction output of \textit{DeepLog} can be extend to detect abnormal sessions of logs containing abnormal behaviours. As reported in \cite{deeplog}, \textit{DeepLog} is first trained using benign sessions containing $M$ logs without any security incidents. After that, to detect anomalies, \textit{DeepLog} is deployed over an input $M$-log long sequence of system logs. Using the first M-1 logs, \textit{DeepLog} then predicts the $M$-th log and compares the prediction output with the observed log. If they are \textit{inconsistent}, \textit{DeepLog} then tags this sequence as anomaly and vice versa.

\subsection{Assessment Setup}
We summarize the datasets used in the assessment pipeline as below. 
The details about the datasets and assessment settings are provided in Appendix B, due to the space limit. All the experimental implementations are available at \url{https://github.com/codesub/AdvCat}.

\noindent \textbf{Fake news detection datasets.} 
We choose two popular datasets: FakeNewsNet \cite{shu2020fakenewsnet} and LIAR \cite{wang2017liar} in the assessment task. 
\textbf{{FakeNewsNet}} contains 369 fake news and 409 real news crawled from \textit{Politifact} website. 
We choose 544 samples for training, 77 samples as the validation set and 157 samples as the test set. 
For sentence paraphrasing in FakeNewsNet, we introduce SCPN \cite{iyyer2018adversarial} to generate neighboring sentences for each sentence in the news text. 
\textbf{{LIAR}} contains 12,791 short statements collected from \textit{Politifact} website. 
These samples belong to 6 categories: pants on fire, false, barely true, half-true, mostly true, and true. 
We regard the first three categories as false and the latter three categories as true. 
We use 10,240 samples for training and 2,551 samples as the test set. 
For word paraphrasing in {LIAR}, we use Paragram-SL999 \cite{wieting2015paraphrase} to generate word paraphrasing neighbors for each word in the texts.
We use the maximum fraction of modifiable words/sentences in each news sample as the perturbation budget. 
For FakeNewsNet, we set the perturbation budget to be $35\%$ and $45\%$ of the length of the news text. 
For LIAR, we conduct word-level perturbations and set budgets as $10\%$ and $20\%$ of the words in each news samples. We set the attack time limit as 3,600 seconds and 1,000 seconds respectively for FakeNewsNet and LIAR.

\noindent \textbf{\textit{Intrusion detection.}} 
We also choose two benchmark datasets - HDFS and IPS - for our assessment.
\textbf{{HDFS}}
\cite{deeplog} contains Hadoop-based executation sessions over 200 Amazon EC2 nodes. 
Each session contain varied number of MapReduce operation logs. 
Following the setting in \cite{deeplog,deepcase}, our model was trained on 4,855 normal sessions of logs, tested on 553,366 normal sessions and 15,200 abnormal sessions. 
\textbf{{IPS}} \cite{tiresias} contains the security event logs collected from 35k sessions by a major security company's intrusion prevention product. 
Each event represents a network-level (e.g., unauthorized login) or system-level activity (e.g., malware detection) that matches a pre-defined signature. 
We use the security events from 30k sessions recorded on a single day for training and the events from 5k sessions for testing. In practices, an adversary can change/remove the security logs to mislead ML-based detection \cite{Pierazzi2020IEEESP}.
For all our experiments, \textit{AdvCat} hence chooses to modify input logs to \textit{DeepLog} for robustness evaluation.
We set the window size of log prediction to 10, the perturbation budget up to 5 event log modifications, and the limit of the run-time of attack to 60 seconds.

\subsection{Robustness Diagnostic Reports}
In both fake news detection and intrusion detection applications, we evaluate the adversarial robustness of the security-critical detection models following the strategy of ``measurement-by-attack'' \cite{AdvGLUE2021NIPS}. We measure the drop of the detection accuracy of the security-critical detection models, when they are exposed to the adversarial attack equipped in \textit{AdvCat}. More specifically, in Table II to Table VI, we show the percentage decrease / increase of the detection accuracy, F1 and False Positive Rate (FPR) compared to those derived in the adversary-free scenario, noted as DAcc, DF1 and DFPR. Besides, we also unveil the average number of required queries and the average running time cost required by each attack method in Fig.2 to Fig.5. 
These empirical observations help the users of \textit{AdvCat} understand the computational cost of robustness evaluation using the service. 

\noindent \textbf{Various state-of-the-art security-critical detection models are significantly vulnerable to adversarial attacks.} \textit{AdvCat} provides detailed diagnostic reports on the utility performance of different fakenews detection models in Table \ref{tab:fakenewsnet_20} and Table \ref{tab:liar_10} with the attack budget of $35\%$ and $10\%$ of the length of the sample on FakeNewsNet and LIAR. Table \ref{tab:fakenewsnet_20} shows that the different fake news detection methods suffers 40\% to 84\% decrease of the detection accuracy with respect to the accuracy and F1 scores and \textbf{2}-\textbf{10} times of the FPR scores obtained at the adversary-free scenario  FakeNewsNet. On LIAR, Table \ref{tab:liar_10} shows that all the involved state-of-the-art detection methods get 57\% to 91\% decrease of the detection accuracy regarding the accuracy and F1 scores, and up to \textbf{2} times of the FPR scores obtained at the adversary-free scenario. The significant decrease of fakenews detection accuracy and increase of FPR scores indicates complete loss of the utility of the state-of-the-art fakenews detection methods facing the attacks.


For the use case of intrusion detection, \textit{AdvCat} first provides the robustness evaluation report of \textit{the log prediction use} of \textit{DeepLog} in Table \ref{tab:deeplog_HDFS} and Table \ref{tab:deeplog_IPS} with the attack budget of 5 (modifying at most 5 logs in each session) on HDFS and IPS for both Top-1 and Top-9 prediction mode. Table \ref{tab:deeplog_HDFS} and Table \ref{tab:deeplog_IPS} show that \textit{DeepLog} suffers from a complete loss of accuracy on HDFS and IPS with respect to both accuracy and F1 score, and significant drop of AUC scores on both datasets. 

For detecting abnormal system sessions, \textit{DeepLog} slides a 10-log long window over each session of logs (each session contains varying number of logs, range from 100 to 20k logs). To perform anomaly detection, one session of security event logs can be tagged as normal only if all the sliding windows are detected as normal. Otherwise, the session is raised as an anomaly. \textit{AdvCat} reports \textit{DR} and \textit{FPR} of anomaly detection with the input perturbation, as well as the variation magnitudes of \textit{DR} (\textit{DDR}$\downarrow$) and \textit{FPR}(\textit{DFPR}$\uparrow$). In our study, only the HDFS data provide labels for normal and abnormal sessions. We therefore conduct the robustness evaluation of \textit{the anomaly detection use} of \textit{DeepLog} on HDFS only. To demonstrate different attack strength setting, in one session of logs in HDFS dataset, we choose respectively 30\% and 60\% of the 10-log long sliding windows to inject the adversarial modification of logs. \textit{AdvCat} hence produces the diagnostic report showing the loss of detection rate and rising the FPR scores of session-level anomaly detection after attack on HDFS in Table \ref{tab:deeplog_session}. As observed in the table, we can find anomaly detection using \textit{DeepLog} gets 100\% of the detection accuracy loss and 2 times of the FPR scores obtained before being exposed to attacks. This is consistent with the empirical results on the log prediction task, as shown in Table \ref{tab:deeplog_IPS}. The state-of-the-art deep learning-based intrusion detection is prone to small perturbation to the input logs, which can drastically bias the anomaly detection outputs. The results raise a severe concern over the trust of the ML-driven intrusion detection model in practical security-critical applications. 

\begin{table}[t]
\caption{Robustness Assessment Results on FakeNewsNet with Attack budget =35\% and Attack Time = 3600sec, reporting the detection performance/percentage of change after the attack is enforced. The most robust  model assessed by each attack algorithm is in bold.
}
\label{tab:fakenewsnet_20}
\centering
\vspace{-2mm}
\resizebox{\linewidth}{!} {
\begin{tabular}{ccccccccc}

\toprule                                                                                                                                                                                            
\textbf{Attack}           & \multicolumn{1}{c}{\textbf{FEmbed}}                & \multicolumn{1}{c}{\textbf{Model}} & \multicolumn{2}{c}{\textbf{Acc/DAcc$\downarrow$}}   & \multicolumn{2}{c}{\textbf{F1/DF1$\downarrow$}}    & \multicolumn{2}{c}{\textbf{FPR/DFPR$\uparrow$}}   \\ \midrule

\multirow{4}{*}{FSGS} & \multicolumn{1}{c}{\multirow{2}{*}{W2V}}  & \multicolumn{1}{c}{MLP}   & \multicolumn{2}{c}{\textbf{0.55 (-40\%)}}  & \multicolumn{2}{c}{\textbf{0.54 (-41\%)}}  & \multicolumn{2}{c}{\textbf{0.34} (+1030\%)}  \\ 
                      & \multicolumn{1}{c}{}                      & \multicolumn{1}{c}{LSTM}  & \multicolumn{2}{c}{0.43 (-49\%)}  & \multicolumn{2}{c}{0.42 (-51\%)} & \multicolumn{2}{c}{0.48 (+242\%)}  \\ \cline{2-9} 
                      & \multicolumn{1}{c}{\multirow{2}{*}{Bert}} & \multicolumn{1}{c}{MLP}   & \multicolumn{2}{c}{0.13 (-84\%)}  & \multicolumn{2}{c}{0.13 (-84\%)} & \multicolumn{2}{c}{0.87 (+335\%)}  \\ 
                      & \multicolumn{1}{c}{}                      & \multicolumn{1}{c}{LSTM}  & \multicolumn{2}{c}{0.24 (-71\%)}  & \multicolumn{2}{c}{0.24 (-71\%)} & \multicolumn{2}{c}{0.85 (+347\%)} \\ \hline
\multirow{4}{*}{SGS}  & \multicolumn{1}{c}{\multirow{2}{*}{W2V}}  & \multicolumn{1}{c}{MLP}   & \multicolumn{2}{c}{\textbf{0.54 (-41\%)}}  & \multicolumn{2}{c}{\textbf{0.53 (-42\%)}}  & \multicolumn{2}{c}{\textbf{0.35} (+1060\%)}   \\ 
                      & \multicolumn{1}{c}{}                      & \multicolumn{1}{c}{LSTM}  & \multicolumn{2}{c}{0.44 (-48\%)}  & \multicolumn{2}{c}{0.43 (-49\%)}  & \multicolumn{2}{c}{0.46 (+228\%)}  \\ \cline{2-9} 
                      & \multicolumn{1}{c}{\multirow{2}{*}{Bert}} & \multicolumn{1}{c}{MLP}   & \multicolumn{2}{c}{0.13 (-84\%)} & \multicolumn{2}{c}{0.13 (-84\%)}& \multicolumn{2}{c}{0.86 (+330\%)}  \\ 
                      & \multicolumn{1}{c}{}                      & \multicolumn{1}{c}{LSTM}  & \multicolumn{2}{c}{0.25 (-70\%)}  & \multicolumn{2}{c}{0.25 (-70\%)} & \multicolumn{2}{c}{0.84 (+342\%)}  \\ \hline
\multirow{4}{*}{UCBS}  & \multicolumn{1}{c}{\multirow{2}{*}{W2V}}  & \multicolumn{1}{c}{MLP}   & \multicolumn{2}{c}{\textbf{0.57 (-38\%)}}  & \multicolumn{2}{c}{\textbf{0.55 (-40\%)}}  & \multicolumn{2}{c}{\textbf{0.32} (+700\%)} \\ 
                      & \multicolumn{1}{c}{}                      & \multicolumn{1}{c}{LSTM}  & \multicolumn{2}{c}{0.45 (-47\%)}  & \multicolumn{2}{c}{0.43 (-49\%)}  & \multicolumn{2}{c}{0.44 (+667\%)}  \\ \cline{2-9}
                      & \multicolumn{1}{c}{\multirow{2}{*}{Bert}} & \multicolumn{1}{c}{MLP}   & \multicolumn{2}{c}{0.19 (-77\%)}  & \multicolumn{2}{c}{0.19 (-77\%)}  & \multicolumn{2}{c}{0.82 (+310\%)} \\ 
                      & \multicolumn{1}{c}{}                      & \multicolumn{1}{c}{LSTM}  & \multicolumn{2}{c}{0.26 (-69\%)}  & \multicolumn{2}{c}{0.26 (-69\%)}& \multicolumn{2}{c}{0.84 (+342\%)} \\ 
\bottomrule
\end{tabular}
}
\end{table}

\begin{figure}[t!]
\small
\centering
\includegraphics[width=0.24\textwidth]{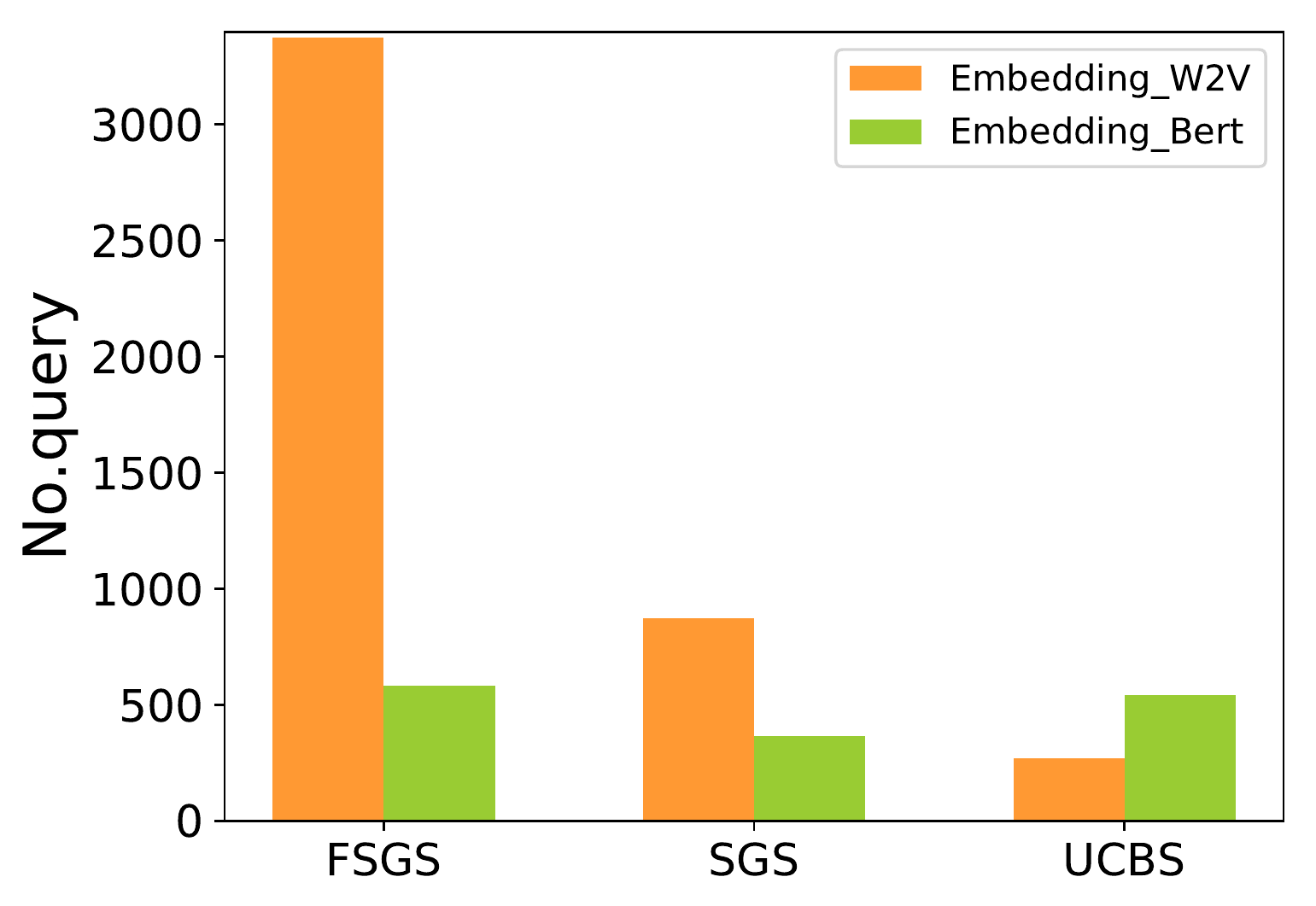}
\includegraphics[width=0.24\textwidth]{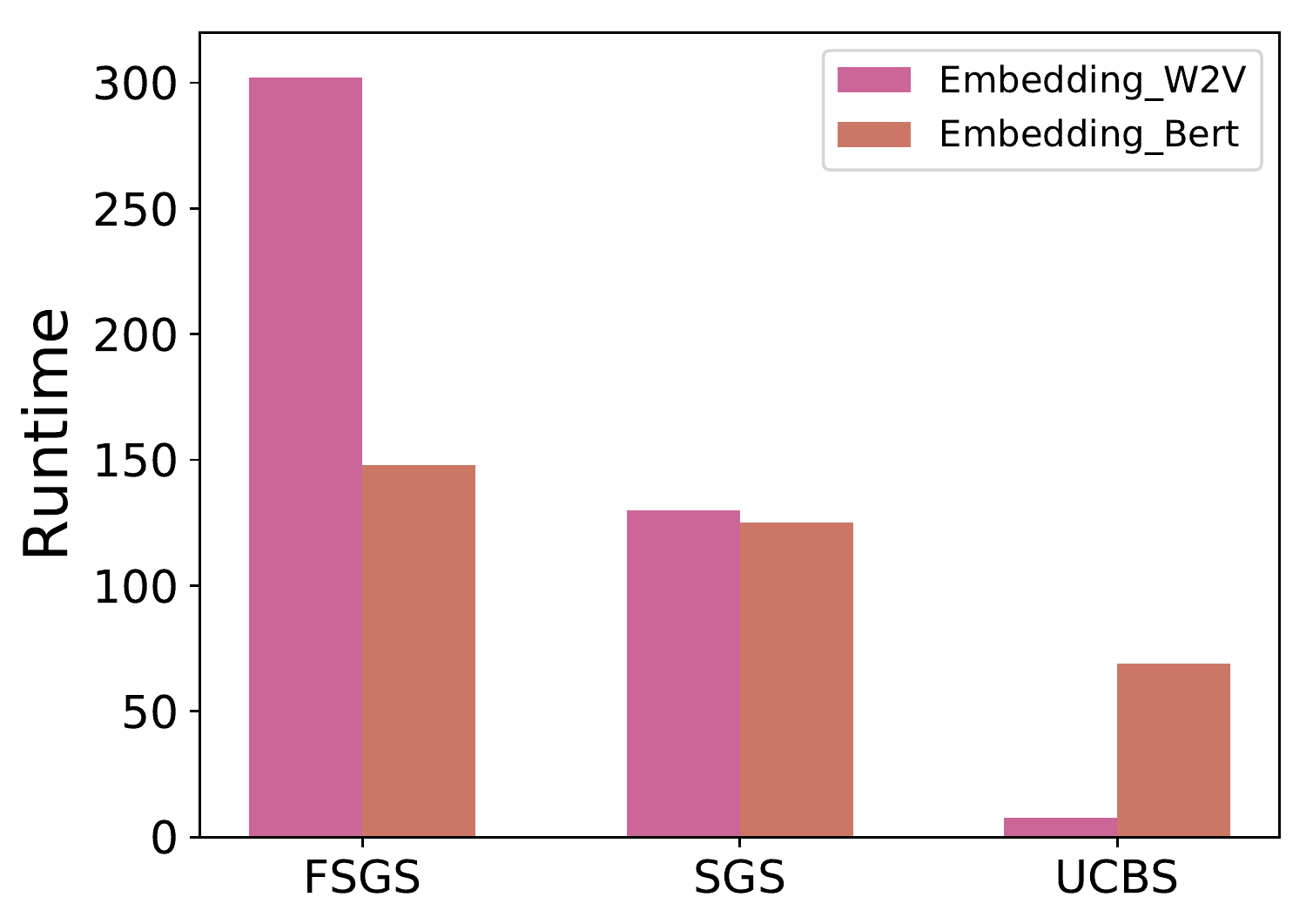}
\vspace{-4mm}
\caption{The assessment expenses on FakeNewsNet dataset. Left: the number of queries. Right: average running time (seconds).}
\label{fig:FakeNews_expense}
\vspace{-0.25cm}
\end{figure}

\begin{table}[t]
\caption{Robustness Assessment Results on LIAR dataset of fake news detection with Attack budget =10\% and Attack Time = 1000sec. The most robust  model assessed by each attack algorithm is in bold.}
\label{tab:liar_10}
\small
\centering
\vspace{-2mm}
\resizebox{0.98\linewidth}{!} {
\begin{tabular}{cccccc}
\toprule                    
\textbf{Attack}           & \multicolumn{1}{c}{\textbf{FEmbed}}                & \multicolumn{1}{c}{\textbf{Model}} & \multicolumn{1}{c}{\textbf{Acc/DAcc$\downarrow$}}   & \multicolumn{1}{c}{\textbf{F1/DF1$\downarrow$}}    & \multicolumn{1}{c}{\textbf{FPR/DFPR$\uparrow$}}  \\ \midrule

\multirow{6}{*}{FSGS}            & \multirow{3}{*}{Glove} & MLP   & 0.07 (-88\%)           & 0.06 (-84\%)          & 0.91 (+139\%)           \\
                                 &                        & LSTM  & 0.11 (-81\%)            & 0.15 (-71\%)          & 0.92 (+149\%)          \\
                                 &                        & CNN   & 0.08 (-87\%)            & 0.05 (-91\%)          & 0.87 (+200\%)          \\ \cline{2-6} 
                                 & \multirow{3}{*}{Bert}  & MLP   & 0.14 (-76\%)            & 0.15 (-76\%)          & \textbf{0.86} (+83\%)            \\
                                 &                        & LSTM  & 0.14 (-77\%)            & 0.18 (-73\%)          & 0.9 (+61\%)           \\
                                 &                        & CNN   & \textbf{0.2 (-67\%)  }           & \textbf{0.27 (-60\%)  }        & 0.89 (+59\%)          \\ \hline
\multirow{6}{*}{SGS}             & \multirow{3}{*}{Glove} & MLP   & 0.06 (-90\%)            & 0.04 (-93\%)          & 0.92 (+142\%)          \\
                                 &                        & LSTM  & 0.1 (-82\%)             & 0.14 (-73\%)          & 0.93 (+151\%)          \\
                                 &                        & CNN   & 0.07 (-89\%)            & 0.04 (-93\%)          & 0.89 (+207\%)           \\ \cline{2-6} 
                                 & \multirow{3}{*}{Bert}  & MLP   & 0.11 (-81\%)            & 0.12 (-81\%)          & \textbf{0.88} (+87\%)          \\
                                 &                        & LSTM  & 0.12 (-74\%)            & 0.17 (-74\%)          & 0.93 (+66\%)          \\
                                 &                        & CNN   & \textbf{0.16 (-74\%)}            & \textbf{0.22 (-68\%)  }        & 0.9 (+61\%)          \\ \hline
\multirow{6}{*}{UCBS}            & \multirow{3}{*}{Glove} & MLP   & 0.12 (-80\%)            & 0.09 (-84\%)          & 0.85 (+124\%)          \\
                                 &                        & LSTM  & 0.17 (-70\%)             & 0.23 (-56\%)          & 0.87 (+135\%)           \\
                                 &                        & CNN   & 0.15 (-76\%)            & 0.07 (-88\%)          & 0.8 (+176\%)           \\ \cline{2-6} 
                                 & \multirow{3}{*}{Bert}  & MLP   & 0.21 (-64\%)           & 0.27 (-57\%)          & 0.83 (+77\%)          \\
                                 &                        & LSTM  & 0.26 (-57\%)            & 0.32 (-52\%)          & \textbf{0.78 (+39\%) }         \\
                                 &                        & CNN   & \textbf{0.33 (-46\%)}            & \textbf{0.42 (-38\%)  }        & 0.79 (+41\%)          \\ \bottomrule
\end{tabular}
}
\vspace{-0.11cm}
\end{table}

\begin{figure}[t!]
\small
\centering
\includegraphics[width=0.245\textwidth]{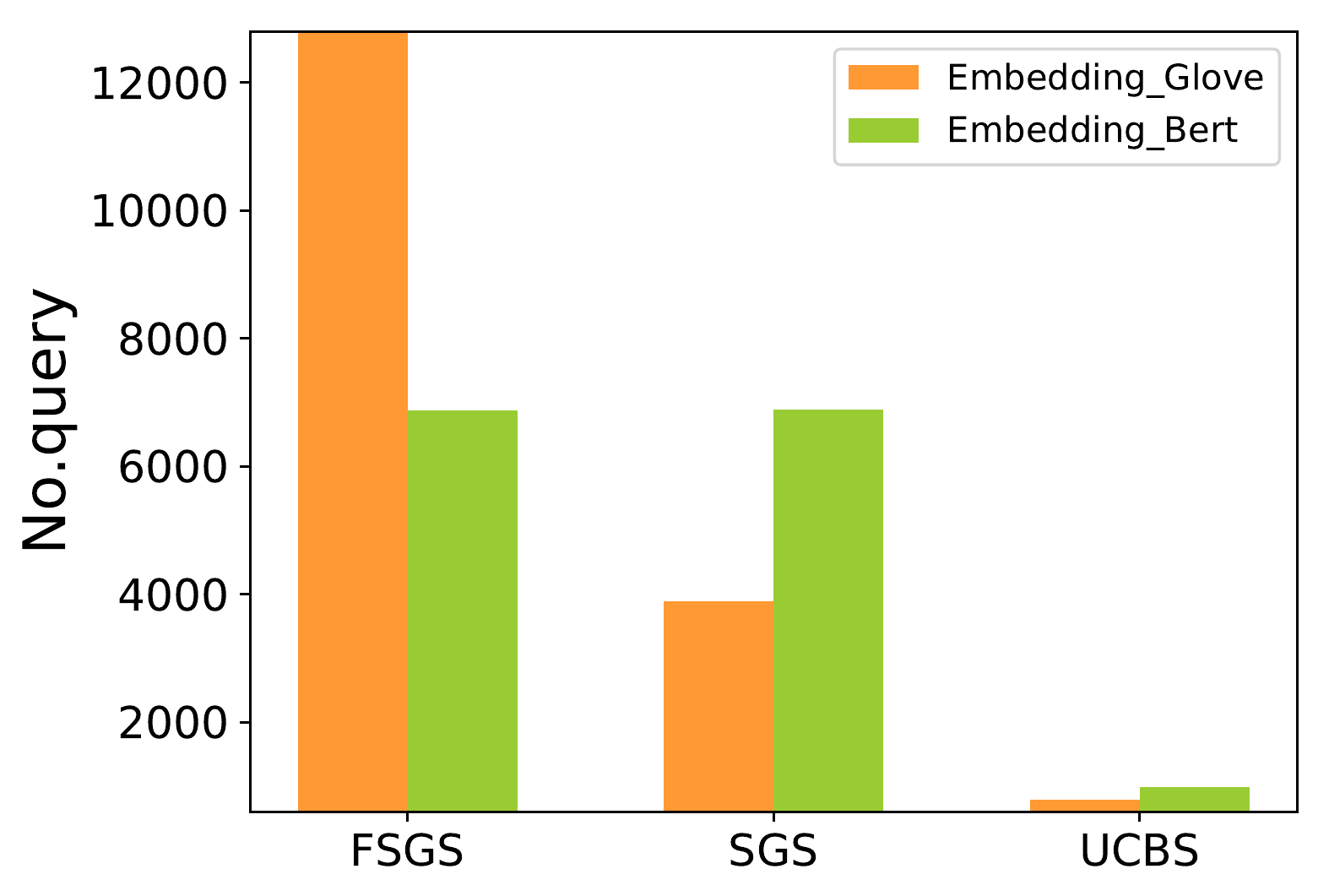}
\includegraphics[width=0.23\textwidth]{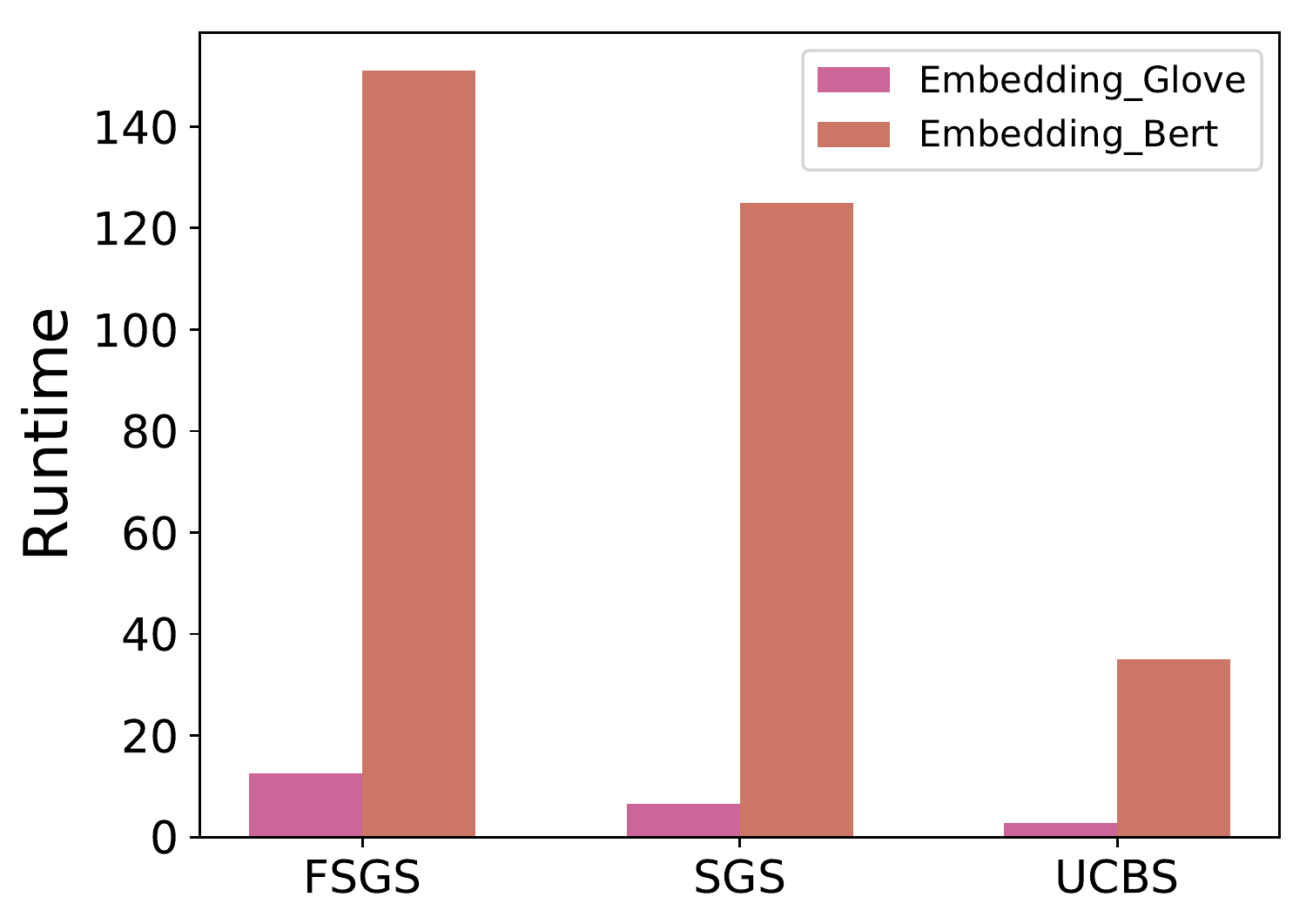}
\vspace{-2mm}
\caption{The assessment expenses on LIAR dataset of fake news detection. Left: the number of queries. Right: average running time (seconds). }
\label{fig:LIAR_FakeNews_expense}
\end{figure}

\noindent\textbf{The most robust models assessed by each attack algorithm.} 
In Table \ref{tab:fakenewsnet_20} and Table \ref{tab:liar_10}, the most robust fakenews detection model assessed by each attack algorithm is highlighted in bold. On the FakeNewsNet dataset, the three attack algorithms have an agreement that the \textbf{MLP with word2vec} is the most robust model. As shown in Table \ref{tab:fakenewsnet_20}, this simple model has the smallest drop of accuracy and F1 score, and has the  highest accuracy and F1 when attack is enforced. It also has the smallest FPR, although a big increase is caused by the attack. This is an interesting assessment result. 

On the other fakenews detection dataset LIAR, all the three attack algorithms
have an agreement that the \textbf{CNN with Bert} is the most robust model, in items of the accuracy/F1 and accuracy/F1 drop. \textbf{MLP with Bert} is reported to have the lowest FPR by FSGS and SGS, while LSTM with Bert is assessed to be the one with the lowest FPR by UCBS.
However, as concluded above, all these models had lost  utility, as indicated by the super low accuracy and high FPR.  \textbf{CNN with Bert} is perhaps recommended as the best candidate model if robustness improvement strategies can be applied. 

\begin{table*}[t]
\caption{Session-level Robustness Assessment of \textit{DeepLog} on HDFS dataset with varied perturbed log-windows.}
\label{tab:deeplog_session}
\small
\centering
\vspace{-2mm}
\resizebox{0.69\linewidth}{!} {
\begin{tabular}{ccccccccccc}
\toprule
 \multirow{2}{*}{\textbf{Method}} &  \multirow{2}{*}{\textbf{Attack}} & \multirow{2}{*}{\textbf{Window}}                   & \multicolumn{4}{c}{\textbf{Top-1}}                        & \multicolumn{4}{c}{\textbf{Top-9}}                                         \\ \cmidrule(r){4-7} \cmidrule(l){8-11} 
    &   &                                                                                  & \multicolumn{2}{c}{\textbf{FPR/DFPR$\uparrow$}}  & \multicolumn{2}{c}{\textbf{DR/DDR$\downarrow$}}       & \multicolumn{2}{c}{\textbf{FPR/DFPR$\uparrow$}}  & \multicolumn{2}{c}{\textbf{DR/DDR$\downarrow$}}      \\ \hline
\multirow{6}{*}{Deeplog}  & \multirow{2}{*}{FSGS} & 60\% window & \multicolumn{2}{c}{1.00 (+212\%)} & \multicolumn{2}{c}{0.00 (-100\%)}   & \multicolumn{2}{c}{1.00 (+99\%)} & \multicolumn{2}{c}{0.00 (-100\%)}   \\ \cline{3-11} 
                          &                       & 30\% window & \multicolumn{2}{c}{1.00 (+212\%)} & \multicolumn{2}{c}{0.00 (-100\%)}   & \multicolumn{2}{c}{1.00 (+99\%)} & \multicolumn{2}{c}{0.00 (-100\%)}   \\ \cline{2-11} 
                          & \multirow{2}{*}{SGS}  & 60\% window & \multicolumn{2}{c}{1.00 (+212\%)} & \multicolumn{2}{c}{0.00 (-100\%)}   & \multicolumn{2}{c}{0.99 (+100\%)} & \multicolumn{2}{c}{0.00 (-100\%)}   \\ \cline{3-11} 
                          &                       & 30\% window & \multicolumn{2}{c}{1.00 (+212\%)} & \multicolumn{2}{c}{0.00 (-100\%)}   & \multicolumn{2}{c}{0.99 (+100\%)} & \multicolumn{2}{c}{0.00 (-100\%)}   \\ \cline{2-11} 
                          & \multirow{2}{*}{UCBS} & 60\% window & \multicolumn{2}{c}{0.99 (+219\%)} & \multicolumn{2}{c}{0.0006 (-100\%)} & \multicolumn{2}{c}{0.97 (+96\%)} & \multicolumn{2}{c}{0.0028 (-100\%)} \\ \cline{3-11} 
                          &                       & 30\% window & \multicolumn{2}{c}{0.99 (+219\%)} & \multicolumn{2}{c}{0.0007 (-100\%)} & \multicolumn{2}{c}{0.94 (+93\%)} & \multicolumn{2}{c}{0.026 (-95\%)}  \\ \bottomrule
\end{tabular}
}
\vspace{-2mm}
\end{table*}

\begin{table}[t]
\caption{Log-window level Robustness Assessment Results of \textit{DeepLog} on HDFS with Attack Budget = $5$ and Attack Time = 60s.}
\label{tab:deeplog_HDFS}
\centering
\vspace{-2mm}
\resizebox{0.95\linewidth}{!} {
\begin{tabular}{cccccccc}

\toprule
 \textbf{Attack} &\textbf{Top-k}  & \multicolumn{2}{c}{\textbf{Acc/DAcc$\downarrow$}}    & \multicolumn{2}{c}{\textbf{F1/DF1$\downarrow$}}      & \multicolumn{2}{c}{\textbf{AUC/DAUC$\downarrow$}}  \\ \hline
\multirow{2}{*}{FSGS} & Top-1   & \multicolumn{2}{c}{0.00 (-100\%)}   & \multicolumn{2}{c}{0.00 (-100\%)}   & \multicolumn{2}{c}{0.53 (-46\%)} \\ \cline{2-8}
                       & Top-9   & \multicolumn{2}{c}{0.00 (-100\%)}   & \multicolumn{2}{c}{0.00 (-100\%)}   & \multicolumn{2}{c}{0.19 (-80\%)} \\ \cline{1-8} 
\multirow{2}{*}{SGS}  & Top-1    & \multicolumn{2}{c}{0.00 (-100\%)}   & \multicolumn{2}{c}{0.00 (-100\%)}   & \multicolumn{2}{c}{0.59 (-40\%)}\\ \cline{2-8}
                       & Top-9   & \multicolumn{2}{c}{0.085 (-91\%)}  & \multicolumn{2}{c}{0.15 (-85\%)}   & \multicolumn{2}{c}{0.19 (-80\%)} \\ \cline{1-8} 
\multirow{2}{*}{UCBS}  & Top-1    & \multicolumn{2}{c}{0.10 (-89\%)}   & \multicolumn{2}{c}{0.18 (-81\%)}   & \multicolumn{2}{c}{0.57 (-42\%)}\\ \cline{2-8}
                       & Top-9   & \multicolumn{2}{c}{0.36 (-64\%\%))}  & \multicolumn{2}{c}{0.53 (-46\%)}   & \multicolumn{2}{c}{0.27 (-73\%)} \\ \bottomrule
\end{tabular}
}
\end{table}

\begin{figure}[t]
\small
\centering
\includegraphics[width=0.243\textwidth]{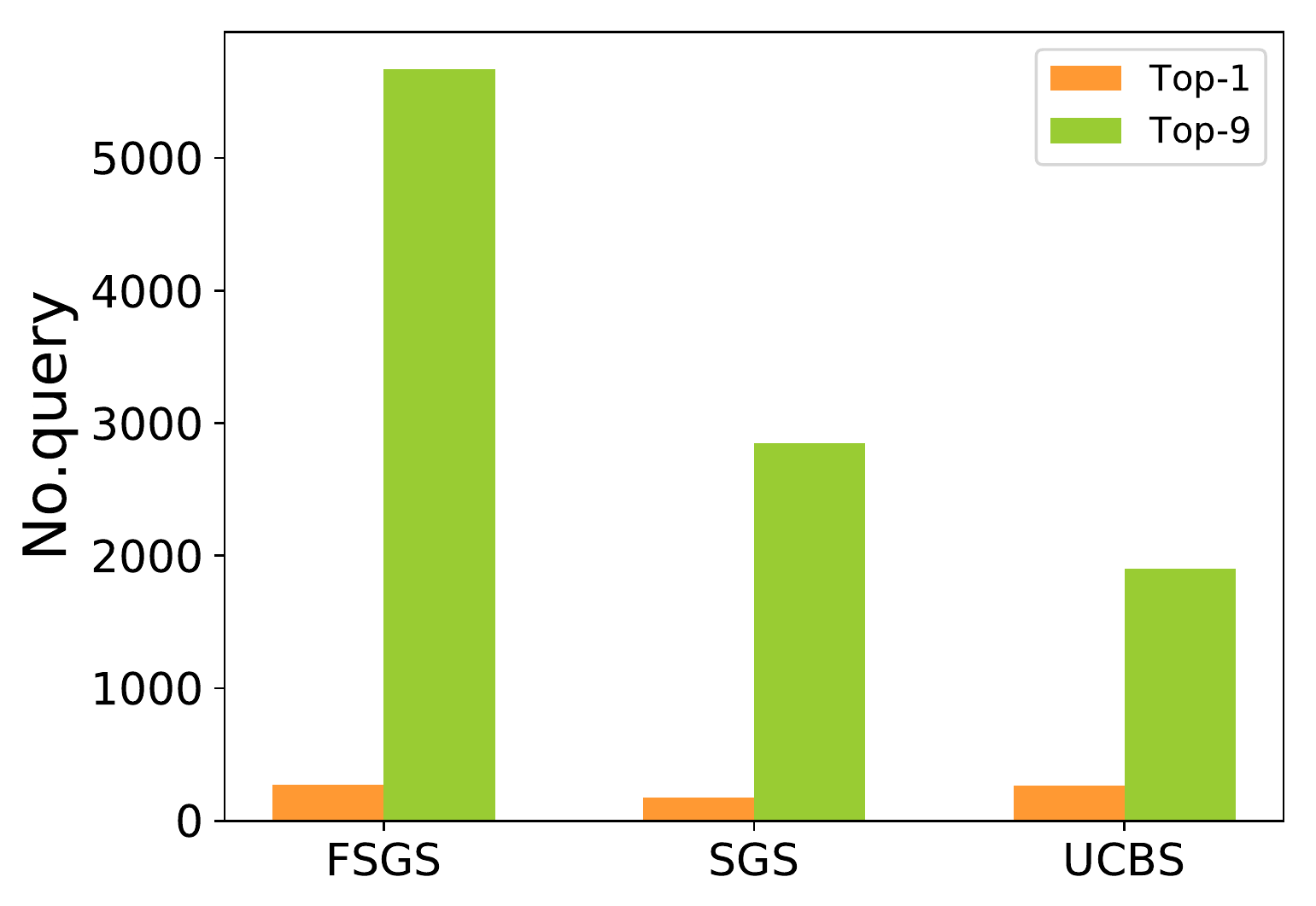}
\includegraphics[width=0.238\textwidth]{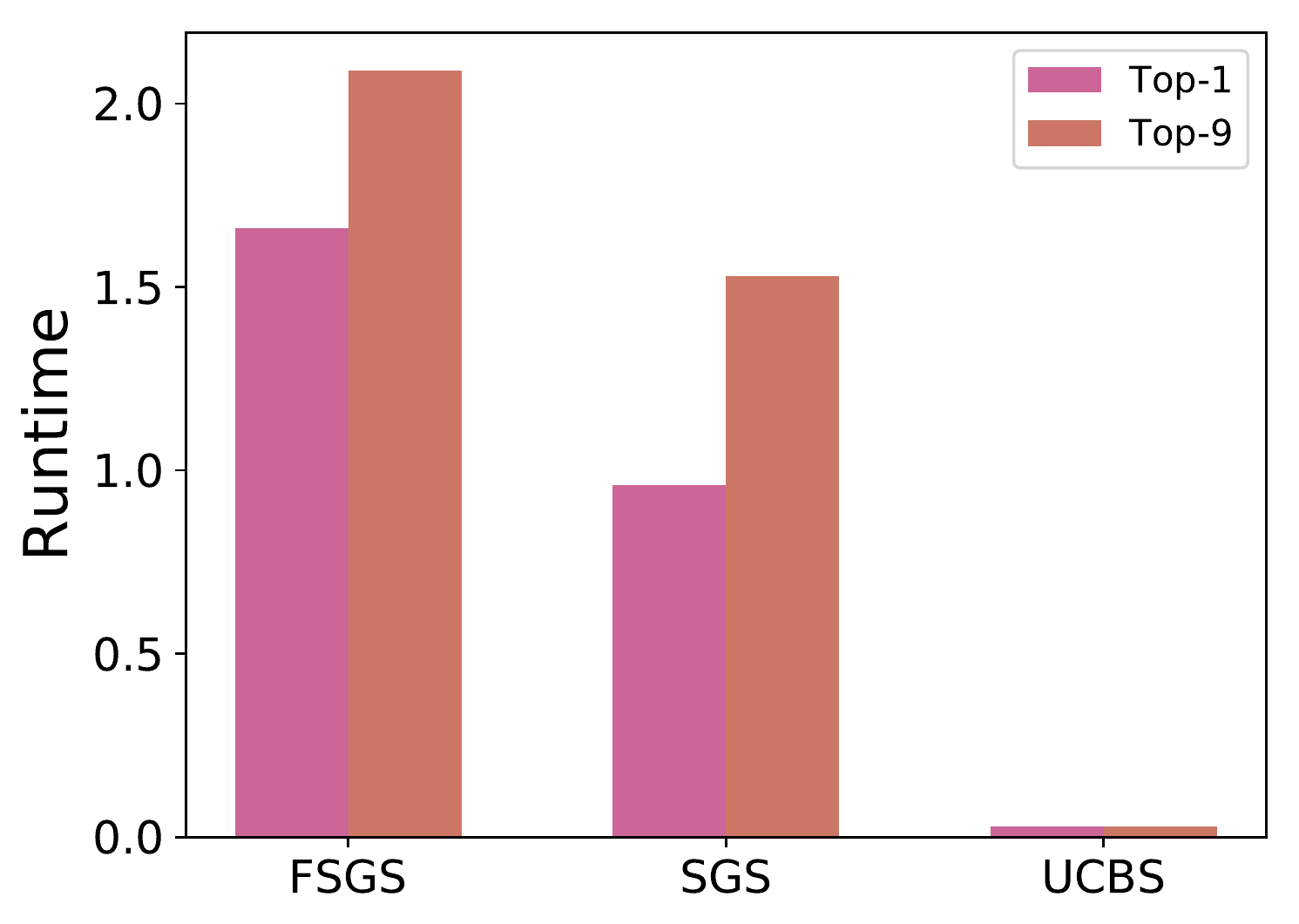}
\vspace{-4mm}
\caption{The assessment expenses on HDFS dataset of intrusion detection. Left: the number of queries. Right: average running time (seconds). }
\label{fig:HDFS_FakeNews_expense}
\vspace{-4mm}
\end{figure}

\begin{table}[t]
\caption{Log-window level Robustness Assessment Results of \textit{DeepLog} on IPS with Attack Budget = $5$ and Attack Time = 60s.}
\label{tab:deeplog_IPS}
\centering
\vspace{-2mm}
\resizebox{0.95\linewidth}{!} {
\begin{tabular}{ccccccccc}
\toprule
\textbf{Attack} &\textbf{Top-k}  & \multicolumn{2}{c}{\textbf{Acc/DAcc$\downarrow$}}    & \multicolumn{2}{c}{\textbf{F1/DF1$\downarrow$}}      & \multicolumn{2}{c}{\textbf{AUC/DAUC$\downarrow$}}  \\ \hline
\multirow{2}{*}{FSGS} & Top-1   & \multicolumn{2}{c}{0.00 (-100\%)}   & \multicolumn{2}{c}{0.00 (-100\%)}   & \multicolumn{2}{c}{0.48 (-44\%)} \\ \cline{2-8}
                      & Top-9   & \multicolumn{2}{c}{0.0024 (-100\%)}& \multicolumn{2}{c}{0.0049 (-100\%)} & \multicolumn{2}{c}{0.47 (-47\%)} \\ \cline{1-8} 
\multirow{2}{*}{SGS}  & Top-1    & \multicolumn{2}{c}{0.00 (-100\%)}   & \multicolumn{2}{c}{0.00 (-100\%)}   & \multicolumn{2}{c}{0.59 (-40\%)}\\ \cline{2-8}
                       & Top-9   & \multicolumn{2}{c}{0.0050 (-100\%)}  & \multicolumn{2}{c}{0.0099 (-100\%)} & \multicolumn{2}{c}{0.50 (-43\%)} \\ \cline{1-8} 
\multirow{2}{*}{UCBS}  & Top-1    & \multicolumn{2}{c}{0.00 (-100\%)}    & \multicolumn{2}{c}{0.00 (-100\%)}   & \multicolumn{2}{c}{0.58 (-32\%)}\\ \cline{2-8}
                       & Top-9   & \multicolumn{2}{c}{0.0051 (-100\%)} & \multicolumn{2}{c}{0.010 (-99\%)}  & \multicolumn{2}{c}{0.42 (-52\%)} \\ \bottomrule 
\end{tabular}
}
\end{table}

\begin{figure}[t]
\small
\centering
\includegraphics[width=0.245\textwidth]{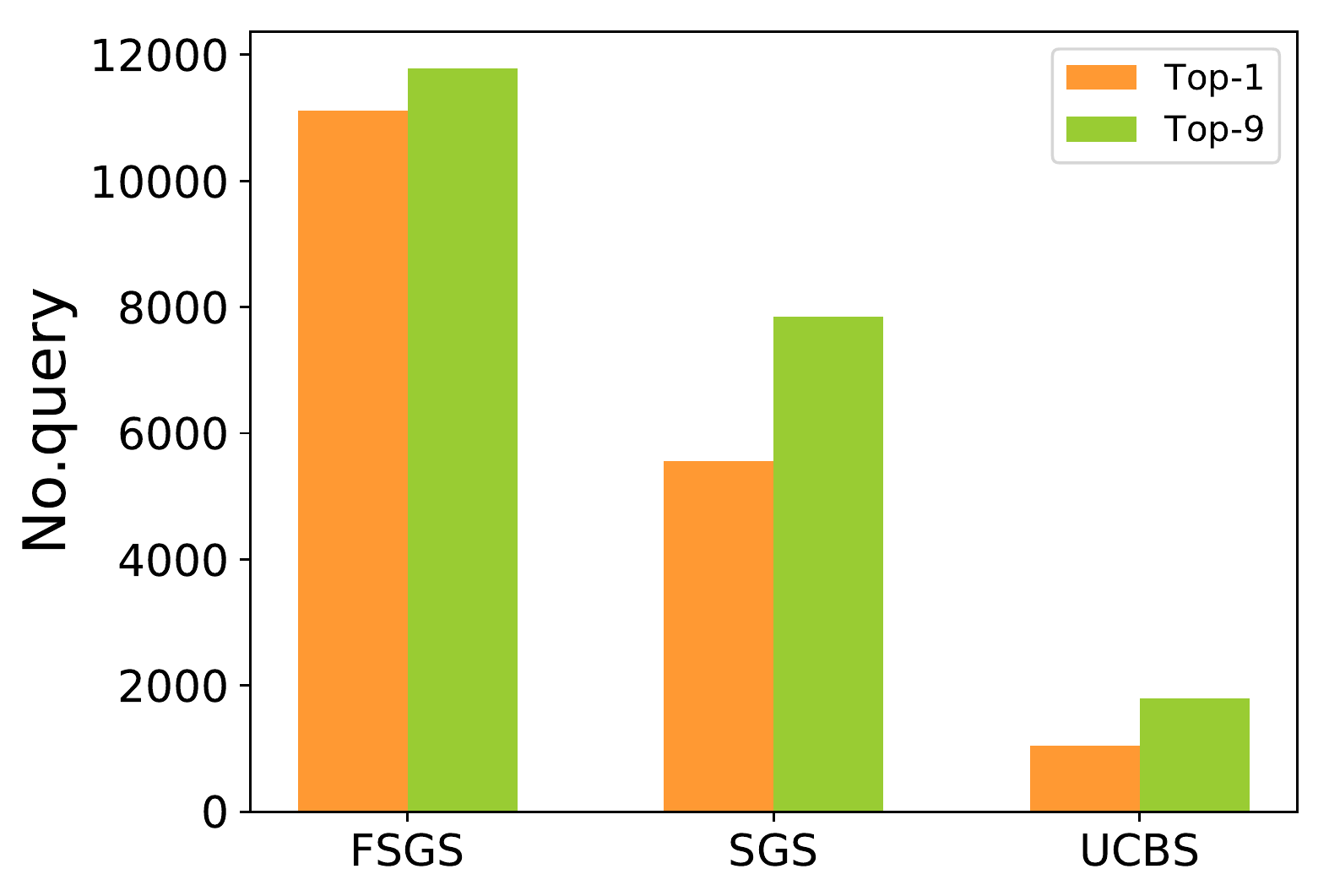}
\includegraphics[width=0.23\textwidth]{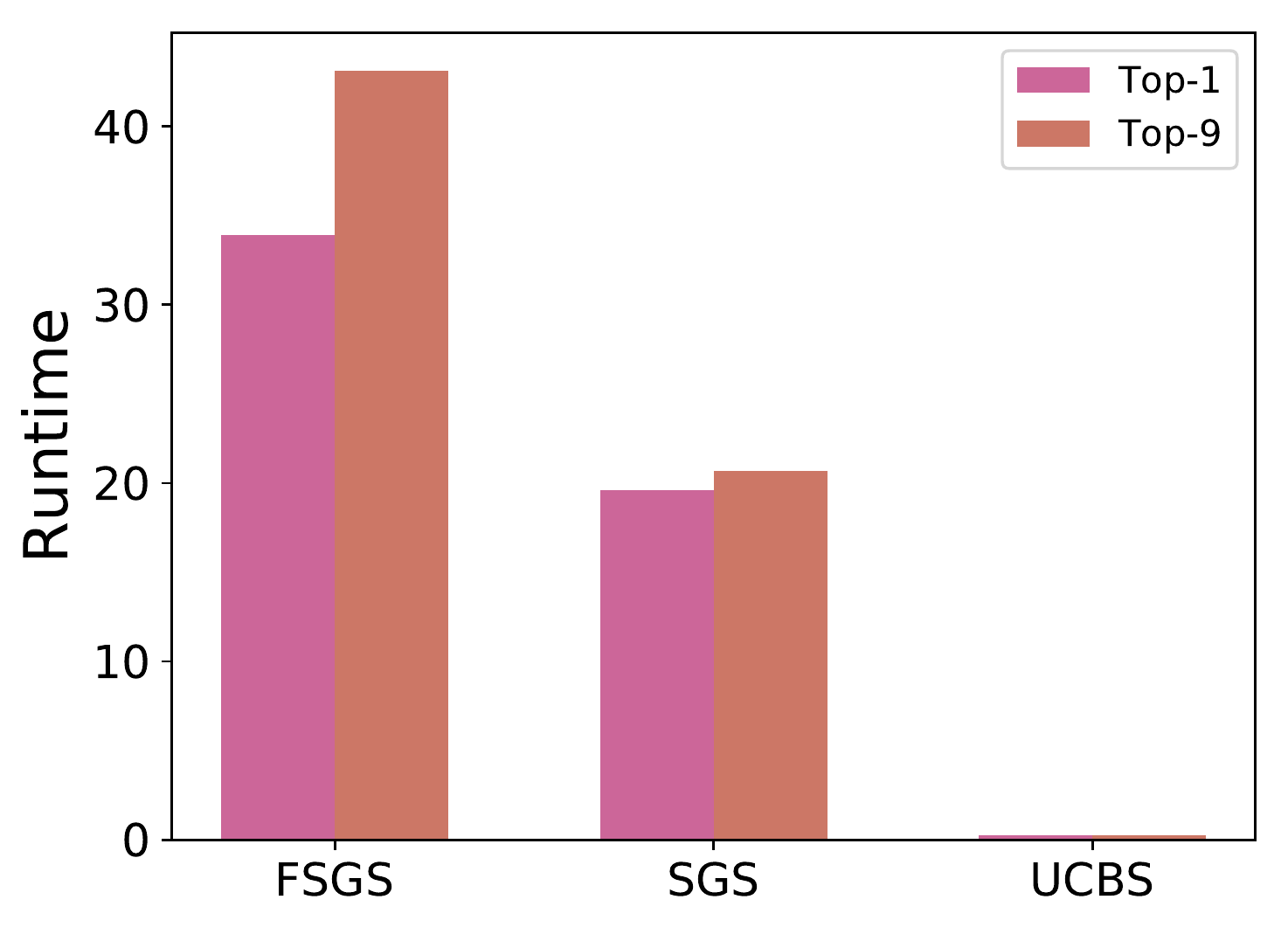}
\vspace{-1mm}
\caption{The assessment expenses on IPS dataset of intrusion detection. Left: the number of queries. Right: average running time (seconds). }
\label{fig:IPS_FakeNews_expense}
\end{figure}

The intrusion detection models have been assessed to have 100\% loss of detection accuracy, as shown in Table \ref{tab:deeplog_HDFS}, \ref{tab:deeplog_IPS} and \ref{tab:deeplog_session}. Their robustness  should be  all improved.  


\noindent\textbf{Computation cost measurement.} On FakeNewsNet, we give the attack cost on the average number of queries and average running time  with word2vec/Bert embedding and MLP/LSTM classifiers in Fig. \ref{fig:FakeNews_expense}. On LIAR, we show the attack cost in the same measures  regarding Glove/Bert embedding, equipped with MLP, LSTM and CNN-based classifiers in Fig. \ref{fig:LIAR_FakeNews_expense}. {It is interesting to find that attacking the Bert embedding-based models cost less queries than the word2vec or Glove embedding based detection models. However, the running-time cost of attacking Bert is higher than the other two embedding techniques. For one thing, as reported in \cite{JinBert2020}, Bert-based embedding is highly prone to the synonym-based adversarial attack methods in text classification, which can be the potential reason behind the observed low query cost. For the other thing, modifying one original word during attack requires to compute the Bert embedding of the synonym substituting the original word. Comparing to word2vec and Glove, computing the Bert embedding requires more time, which explains the relatively higher running time of attacking Bert-based models. } 

For the intrusion detection use cases, on HDFS and IPS, we demonstrate the attack cost on the average of the average number of query and average running time in Top-1 and Top-9 based models from Fig. \ref{fig:HDFS_FakeNews_expense} and Fig. \ref{fig:IPS_FakeNews_expense}. 
Success of the Top-9 prediction only requires the true log-to-predict exists in the Top-9 ranked outputs according to the prediction confidence. In contrast, the Top-1 prediction requires the true log-to-predict has the highest prediction confidence. Therefore, Top-1 is less stable facing adversarial perturbations. As unveiled, we find that attacking the Top-1 mode prediction models on both HDFS and IPS needs less average time and query number than the Top-9 mode.
As observed in Fig. \ref{fig:FakeNews_expense} to Fig. \ref{fig:IPS_FakeNews_expense}, we show a trade-off between attack effectivenes and computational efficiency existing in the three attack methods offered by \textit{AdvCat}. Among all the three attack methods, on one hand, UCBS and FSGS spend consistently the least and most average running time and query cost respectively. SGS ranks in between them. UCBS shows the order of magnitudes higher efficiency than FSGS in perturbing the target detection model. On the other hand, across different detection models and security-critical use cases, FSGS can achieve more stronger attack effects, causing significantly lower detection accuracy and F1 score and higher FPR values than SGS and UCBS. The users of \textit{AdvCat} can choose the proper attack method upon their  budgets of computational resources for practical robustness evaluation tasks. In general, the experimental results show that \textit{AdvCat} can provide the robustness alert to potential vulnerable detection models before being deployed in practices. 

\section{Conclusion}
\textit{AdvCat} is established as a domain- and model-agnostic pipeline to evaluate the adversarial robustness of ML-driven cybersecurity-critical applications with categorical features. We address the challenges from two perspectives. First, we propose to integrate three domain-agnostic assessment methods into \textit{AdvCat}. Based on them, \textit{AdvCat} can provide assessment-as-a-service to any ML-driven cybersecurity applications, without accessing the target model and privacy-sensitive contents of the application data. Second, we build the quality and computational cost analysis of these methods, which guarantee the use of the assessment pipeline for general classifiers. Via substantial empirical study, we demonstrate the use of \textit{AdvCat} over fake news and intrusion detection use cases. The results unveil the state-of-the-art ML-driven cybersecurity applications are highly vulnerable to adversarial attacks. In future, we plan to extend the use of \textit{AdvCat} to more real-world cybersecurity use cases. 
\section*{Acknowledgements}
The research reported in this paper was partially supported by funding from King Abdullah University of Science and Technology (KAUST).
\bibliographystyle{IEEEtran}
\bibliography{reference_simplified}

\clearpage
\appendix
\section{Perturbation-Free Performances of Fake News and Intrusion Detection}
We show the fake news detection performances including \textit{Accuracy}, \textit{F1 Score}, \textit{AUC Score} or \textit{False Positive Rate}. We evaluated these performance before and after attack and use \textit{DAcc}, \textit{DF1}, \textit{DAUC} or \textit{DFPR} respectively to denote the descending value or ascending value after attack. In Table.\ref{tab:model_baseline}, we provide the perturbation-free performances of different fake news detection models over the two datasets. They are obtained without injecting adversarial perturbations, which measures the adversary-free utility baseline of fake news detection. 

We summarize the adversary-free performances of \textit{DeepLog} as in Table.\ref{tab:intrusion_baseline}. We measure \textit{Acc}, \textit{F1}, \textit{FPR} or \textit{AUC} of the log-prediction results. For session-level anomaly detection on the HDFS data, \textit{DR} and \textit{FPR} of the abnormal session detection result using \textit{DeepLog} are 0.63 and 0.31 for the top-1 prediction mode, and 0.55 and 0.0013 for the top-9 prediction mode. 
\begin{table}[h]
\small
\caption{Perturbation-Free Fake News Detection Accuracy}
\label{tab:model_baseline}
\setbox0\hbox{\tabular{@{}l}\textbf{FakeNewsNet}\endtabular}
\setlength{\extrarowheight}{2.2pt}
\resizebox{\linewidth}{!} {
\begin{tabular}{c|cccccccc}
\toprule
\textbf{Dataset}  & \textbf{FEm}    & \textbf{Model} &  \textbf{ACC}  & \textbf{Pre} & \textbf{Rec} & \textbf{F1}   & \textbf{AUC}  & \textbf{FPR}  \\ \midrule
\multirow{4}{*}[1.3ex]{\rotatebox{90}{\usebox0}} & \multirow{2}{*}{W2V}          & MLP                             & 0.92                          & 0.95                          & 0.86                          & 0.92                         & 0.96                          & 0.04                          \\
                                  &                               & LSTM                            & 0.85                          & 0.84                          & 0.85                          & 0.85                         & 0.90                          & 0.14                          \\ \cline{2-9} 
                                  & \multirow{2}{*}{Bert}         & MLP                             & 0.82                          & 0.78                          & 0.84                          & 0.82                         & 0.88                          & 0.20                          \\
                                  &                               & LSTM                            & 0.83                          & 0.81                          & 0.85                          & 0.82                         & 0.88                          & 0.19  \\         \midrule  
\multirow{6}{*}{\rotatebox[origin=c]{90}{\textbf{LIAR}}} & \multirow{3}{*}{Glove}        & MLP                             & 0.59                          & 0.55                          & 0.55                          & 0.55                         & 0.50                          & 0.38                          \\
                                  &                               & LSTM                            & 0.57                          & 0.54                          & 0.50                          & 0.52                         & 0.47                          & 0.37                          \\
                                  &                               & CNN                             & 0.62                          & 0.60                          & 0.52                          & 0.56                         & 0.46                          & 0.29                          \\ \cline{2-9} 
                                  & \multirow{3}{*}{Bert}         & MLP                             & 0.59                          & 0.62                          & 0.65                          & 0.63                         & 0.53                          & 0.47                          \\
                                  &                               & LSTM                            & 0.60                          & 0.60                          & 0.74                          & 0.66                         & 0.55                          & 0.56                          \\
                                  &                               & CNN                             & 0.61                          & 0.61                          & 0.76                          & 0.68                         & 0.56                          & 0.56                          \\ \bottomrule
\end{tabular}
}
\end{table}



\begin{table}[t]
\small
\caption{Perturbation-Free Log-window Level Detection Performance of Intrusion Detection}
\label{tab:intrusion_baseline}
\setbox0\hbox{\tabular{@{}l}\textbf{IPS}\endtabular}
\setbox1\hbox{\tabular{@{}l}\textbf{HDFS}\endtabular}
\resizebox{\linewidth}{!} {
\begin{tabular}{c|ccccccc}
\toprule
\textbf{Dataset}  & \textbf{Top-k}    & \textbf{Model} &  \textbf{ACC}  & \textbf{Pre} & \textbf{Rec} & \textbf{F1}   & \textbf{AUC}     \\ \midrule
\multirow{6}{*}[5ex]{{\usebox1}} & \multirow{1}{*}{Top-1}  & DeepLog   & 0.91 & 1.00 & 0.91 & 0.96 & 0.99  \\ \cline{2-8} 
                      & \multirow{1}{*}{Top-9} & DeepLog   & 0.99 & 1.00 & 1.00 & 1.00 & 0.99 \\ \hline
\multirow{2}{*}[0.1ex]{{\usebox0}}
  & \multirow{1}{*}{Top-1}  & DeepLog   & 0.58 & 1.00 & 0.59 & 0.74 & 0.87 \\ \cline{2-8}
  & \multirow{1}{*}{Top-9} & DeepLog   & 0.95 & 1.00 & 0.96   & 0.98 & 0.87  \\ 
\bottomrule
\end{tabular}
}
\end{table}

\section{Details about Assessment Datasets and Setup}
We conduct the experimental study using PyTorch on a Ubuntu workstation with two NVIDIA 3090 GPUs.

\textbf{{FakeNewsNet}} 
contains news evaluated by exports or journalists and tagged as fake or real news. FakeNewsNet also contains the information about social contexts and spatiotemporal information. \cite{dou2021user} adopts Graph Neural Networks integrating these additional records with news contents and historical posts of the users spreading the news to help classification. In our study, we only focus on the adversarial perturbation over the news contents. 

\textbf{{LIAR}} is also collected from \textit{Politifact} website. Different from {FakeNewsNet}, LIAR contains mostly short statements without meta attributes. It collects 12,791 samples and they are classified into 6 categories: pants on fire, false, barely true, half-true, mostly true, and true. For our task, we regard the first three categories as false and the latter three categories as true. We use 10,240 samples for training and 2551 samples as the test set.

As the news samples of {FakeNewsNet} contain longer texts than those of {LIAR}, we apply sentence-level perturbations for {FakeNewsNet} and word-level perturbations for {LIAR}. For word paraphrasing in {LIAR}, we use Paragram-SL999 \cite{wieting2015paraphrase} to generate word paraphrasing neighbors for each word in the texts. For sentence paraphrasing in {FakeNewsNet}, we introduce SCPN \cite{iyyer2018adversarial} to generate neighboring sentences for each sentence in the news text. Note that we treat the neighboring words/sentences as alternative categorical values of each word/sentence in the original news texts. They define the domain-specific rules to bound the perturbation search space over the news texts. While conducting assessment, \textit{AdvCat} only needs to know the indexes of these neighboring words/sentences. In other words, the testing news samples are anonymized. 

For {FakeNewsNet} and {LIAR}, we set the time limit as 3,600 seconds and 1,000 seconds respectively. As different samples have different numbers of words or sentences, we use the maximum fraction of modifiable words/sentences in each news sample as the perturbation budget. For each dataset, we test with two budget settings to assess the robustness under different degrees of attack. For FakeNewsNet, we conduct sentence-level perturbations. There are over 30\% of the news samples containing less than 3 sentences. Therefore, for each testing news sample in {FakeNewsNet}, we choose the perturbation budget to be $35\%$ and $45\%$ of the length of the news text. For LIAR, we conduct word-level perturbations and set budgets as $10\%$ and $20\%$ of the words in each news samples.

\begin{table}[t]
\caption{Robustness Assessment Results on FakeNewsNet with Attack budget =45\% and Attack Time = 3600s}
\label{tab:fakenewsnet_40}
\small
\resizebox{\linewidth}{!} {
\begin{tabular}{ccccccccc}
\toprule                                                                                                                                                                                  
\textbf{Attack}           & \multicolumn{1}{c}{\textbf{FEmbed}}                & \multicolumn{1}{c}{\textbf{Model}} & \multicolumn{2}{c}{\textbf{Acc/DAcc$\downarrow$}}  & \multicolumn{2}{c}{\textbf{F1/DF1$\downarrow$}}    & \multicolumn{2}{c}{\textbf{FPR/DFPR$\uparrow$}}  \\ \midrule
\multirow{4}{*}{FSGS}            & \multirow{2}{*}{W2V}  & MLP   & \multicolumn{2}{c}{0.46 (-50\%)}            & \multicolumn{2}{c}{0.43 (-53\%)}          & \multicolumn{2}{c}{0.40 (+900\%)}          \\
                                 &                       & LSTM  & \multicolumn{2}{c}{0.35 (-59\%)}            & \multicolumn{2}{c}{0.32 (-62\%)}          & \multicolumn{2}{c}{0.48 (+243\%)}          \\ \cline{2-9} 
                                 & \multirow{2}{*}{Bert} & MLP   & \multicolumn{2}{c}{0.10 (-89\%)}            & \multicolumn{2}{c}{0.10 (-89\%)}          & \multicolumn{2}{c}{0.88 (+340\%)}          \\
                                 &                       & LSTM  & \multicolumn{2}{c}{0.17 (-80\%)}            & \multicolumn{2}{c}{0.17 (-79\%)}          & \multicolumn{2}{c}{0.86 (+353\%)}          \\ \hline
\multirow{4}{*}{SGS}             & \multirow{2}{*}{W2V}  & MLP   & \multicolumn{2}{c}{0.47 (-48\%)}            & \multicolumn{2}{c}{0.44 (-52\%)}          & \multicolumn{2}{c}{0.38 (+850\%)}          \\
                                 &                       & LSTM  & \multicolumn{2}{c}{0.36 (-58\%)}            & \multicolumn{2}{c}{0.33 (-62\%)}          & \multicolumn{2}{c}{0.47 (+236\%)}          \\ \cline{2-9} 
                                 & \multirow{2}{*}{Bert} & MLP   & \multicolumn{2}{c}{0.12 (-85\%)}            & \multicolumn{2}{c}{0.12 (-85\%)}          & \multicolumn{2}{c}{0.84 (+320\%)}          \\
                                 &                       & LSTM  & \multicolumn{2}{c}{0.17 (-80\%)}            & \multicolumn{2}{c}{0.17 (-79\%)}          & \multicolumn{2}{c}{0.86 (+353\%)}          \\ \hline
\multirow{4}{*}{UCBS}            & \multirow{2}{*}{W2V}  & MLP   & \multicolumn{2}{c}{0.45 (-57\%)}            & \multicolumn{2}{c}{0.42 (-53\%)}          & \multicolumn{2}{c}{0.40 (+900\%)}          \\
                                 &                       & LSTM  & \multicolumn{2}{c}{0.36 (-58\%)}            & \multicolumn{2}{c}{0.33 (-61\%)}          & \multicolumn{2}{c}{0.49 (+250\%)}          \\ \cline{2-9} 
                                 & \multirow{2}{*}{Bert} & MLP   & \multicolumn{2}{c}{0.14 (-83\%)}            & \multicolumn{2}{c}{0.14 (-83\%)}          & \multicolumn{2}{c}{0.85 (+325\%)}          \\
                                 &                       & LSTM  & \multicolumn{2}{c}{0.18 (-78\%)}            & \multicolumn{2}{c}{0.17 (-79\%)}          & \multicolumn{2}{c}{0.87 (+358\%)}          \\ \bottomrule
\end{tabular}
}
\end{table}

\begin{figure}[h!]
\small
\centering
\includegraphics[width=0.23\textwidth]{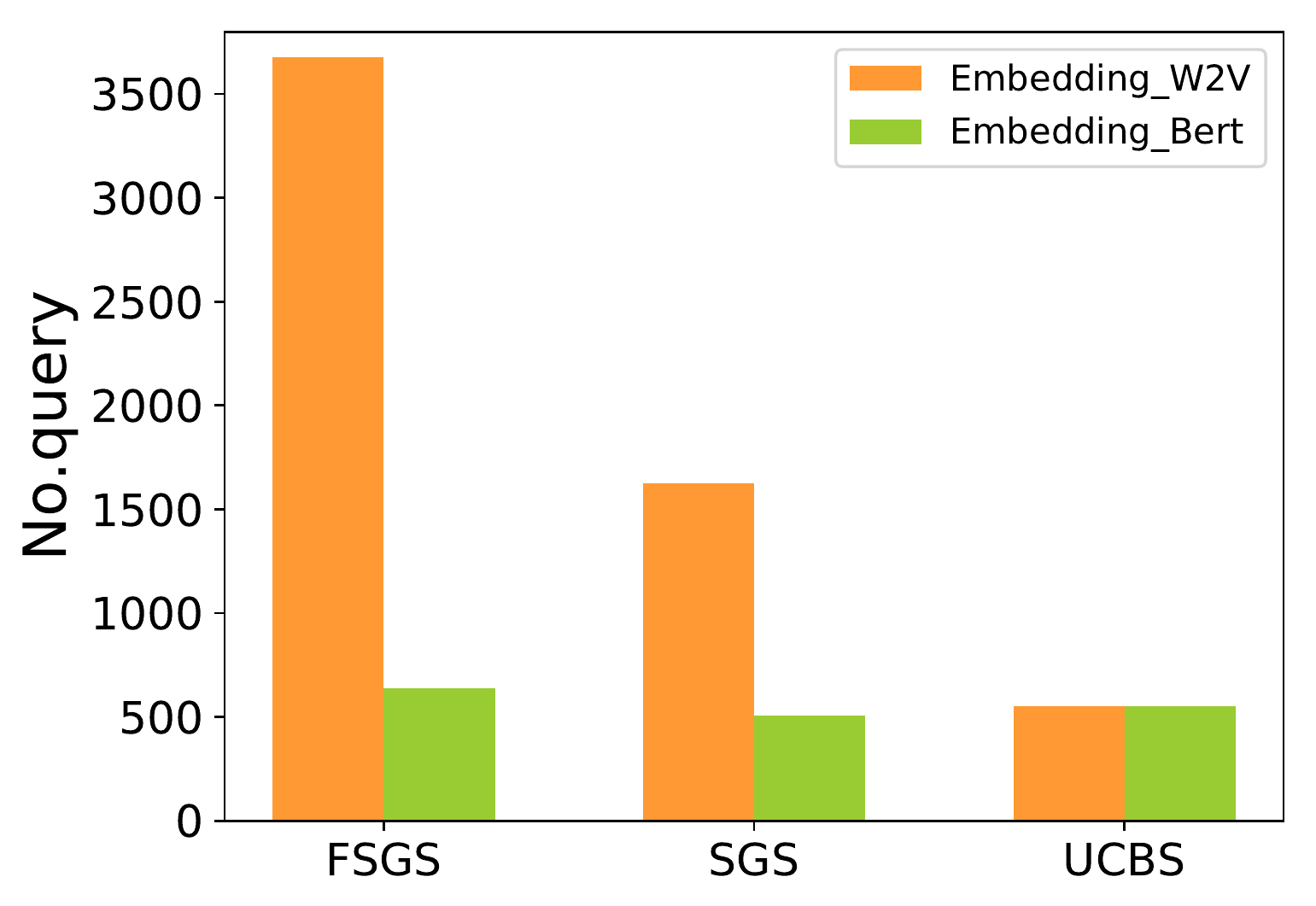}
\includegraphics[width=0.23\textwidth]{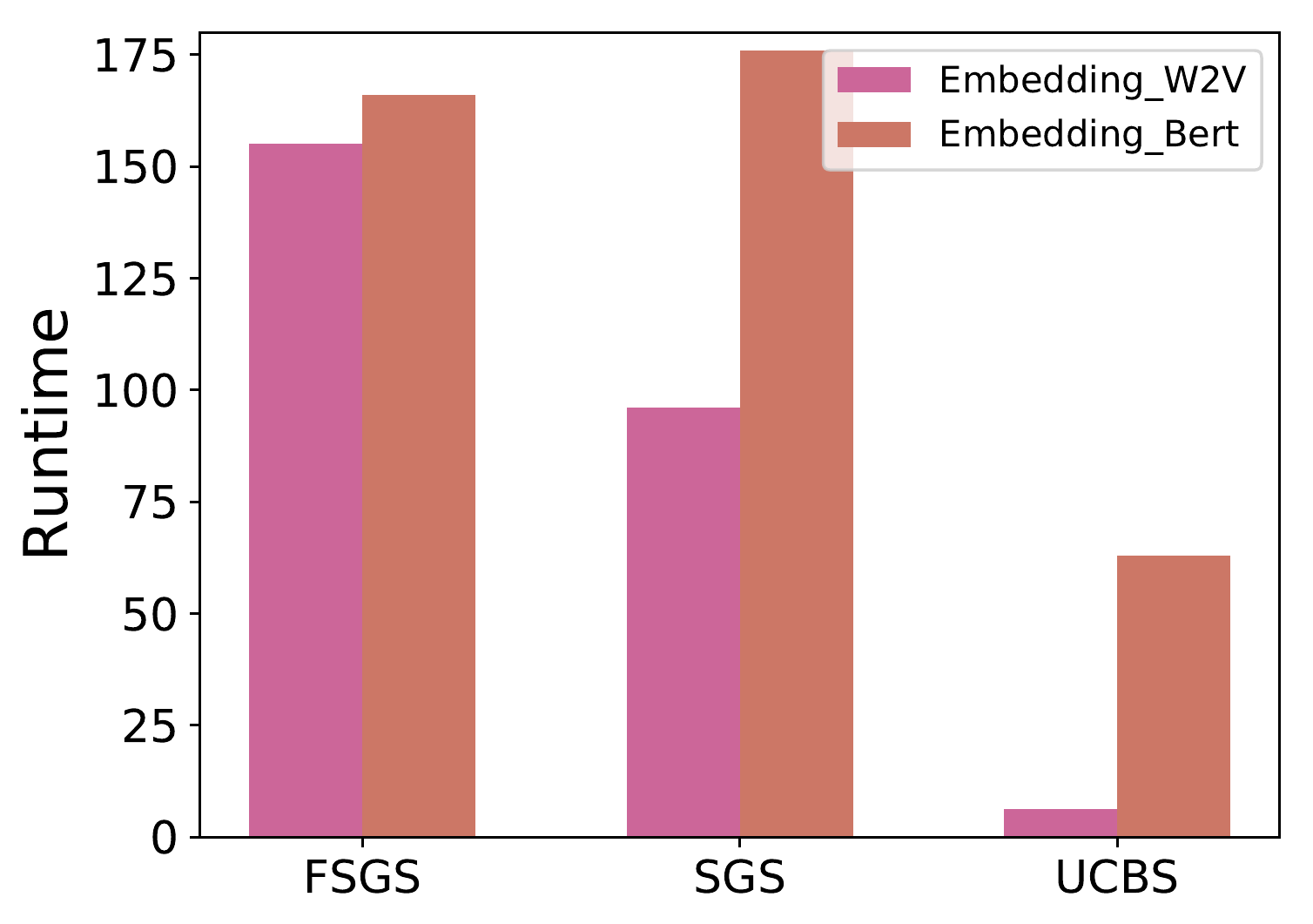}
\vspace{-2mm}
\caption{The assessment expenses on FakeNewsNet dataset of fakenews detection . Left: the number of queries. Right: average running time.}
\label{fig:FakeNews_Runtime_app}
\vspace{-0.1cm}
\end{figure}

\begin{table}[t]
\small
\caption{Robustness Assessment Results on LIAR with Attack budget = 20\% and Attack Time = 1000s.}
\label{tab:liar_20_appendix}
\resizebox{\linewidth}{!} {
\begin{tabular}{cccccc}
\toprule                    
\textbf{Attack}           & \multicolumn{1}{c}{\textbf{FEmbed}}                & \multicolumn{1}{c}{\textbf{Model}} & \multicolumn{1}{c}{\textbf{Acc/DAcc$\downarrow$}}  & \multicolumn{1}{c}{\textbf{F1/DF1$\downarrow$}}    & \multicolumn{1}{c}{\textbf{FPR/DFPR$\uparrow$}}  \\ \midrule
\multirow{6}{*}{FSGS}                             & \multirow{3}{*}{Glove} & MLP   & 0.05 (-92\%)            & 0.04 (-93\%)          & 0.94 (+147\%)          \\
                                                  &                        & LSTM  & 0.1 (-82\%)             & 0.13 (-75\%)          & 0.93 (+151\%)          \\
                                                  &                        & CNN   & 0.06 (-90\%)            & 0.03 (-95\%)          & 0.97 (+234\%)          \\ \cline{2-6} 
                                                  & \multirow{3}{*}{Bert}  & MLP   & 0.11 (-81\%)            & 0.13 (-79\%)           & 0.91 (+94\%)          \\
                                                  &                        & LSTM  & 0.12 (-80\%)            & 0.16 (-76\%)           & 0.92 (+64\%)          \\
                                                  &                        & CNN   & 0.18 (-70\%)            & 0.24 (-65\%)          & 0.9 (+61\%)           \\ \hline
\multirow{6}{*}{SGS}                              & \multirow{3}{*}{Glove} & MLP   & 0.05 (-91\%)            & 0.04 (-93\%)          & 0.95 (+150\%)          \\
                                                  &                        & LSTM  & 0.09 (-84\%)            & 0.13 (-75\%)          & 0.94 (+154\%)          \\
                                                  &                        & CNN   & 0.06 (-90\%)            & 0.03 (-95\%)          & 0.91 (+213\%)          \\ \cline{2-6} 
                                                  & \multirow{3}{*}{Bert}  & MLP   & 0.1 (-83\%)             & 0.11 (-83\%)          & 0.9 (+0.91\%)           \\
                                                  &                        & LSTM  & 0.11 (-82\%)            & 0.15 (-77\%)          & 0.92 (+64\%)          \\
                                                  &                        & CNN   & 0.16 (-74\%)            & 0.23 (-66\%)          & 0.91 (+63\%)          \\ \hline
\multirow{6}{*}{UCBS}                             & \multirow{3}{*}{Glove} & MLP   & 0.09 (-85\%)             & 0.07 (-87\%)          & 0.93 (+145\%)          \\
                                                  &                        & LSTM  & 0.15 (-74\%)            & 0.19 (-63\%)          & 0.87 (+135\%)          \\
                                                  &                        & CNN   & 0.12 (-81\%)             & 0.06 (-89\%)           & 0.85 (+193\%)          \\ \cline{2-6} 
                                                  & \multirow{3}{*}{Bert}  & MLP   & 0.18 (-69\%)            & 0.21 (-67\%)          & 0.82 (+74\%)          \\
                                                  &                        & LSTM  & 0.18 (-70\%)            & 0.23 (-65\%)          & 0.84 (+50\%)          \\
                                                  &                        & CNN   & 0.27 (-56\%)            & 0.36 (-47\%)          & 0.82 (+46\%)          \\ \hline
\end{tabular}
}
\end{table}

\begin{figure}[h]
\small
\centering
\includegraphics[width=0.23\textwidth]{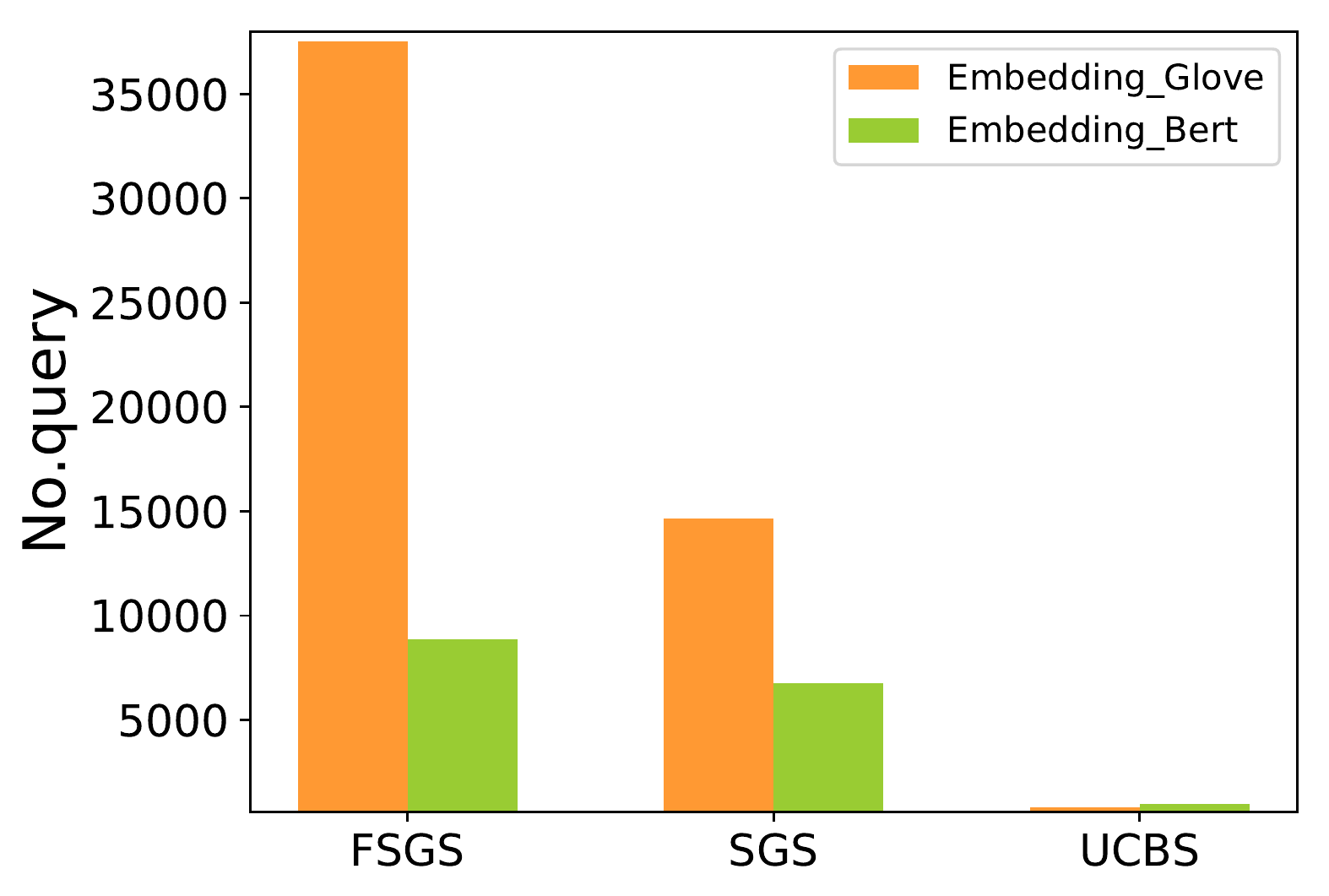}
\includegraphics[width=0.23\textwidth]{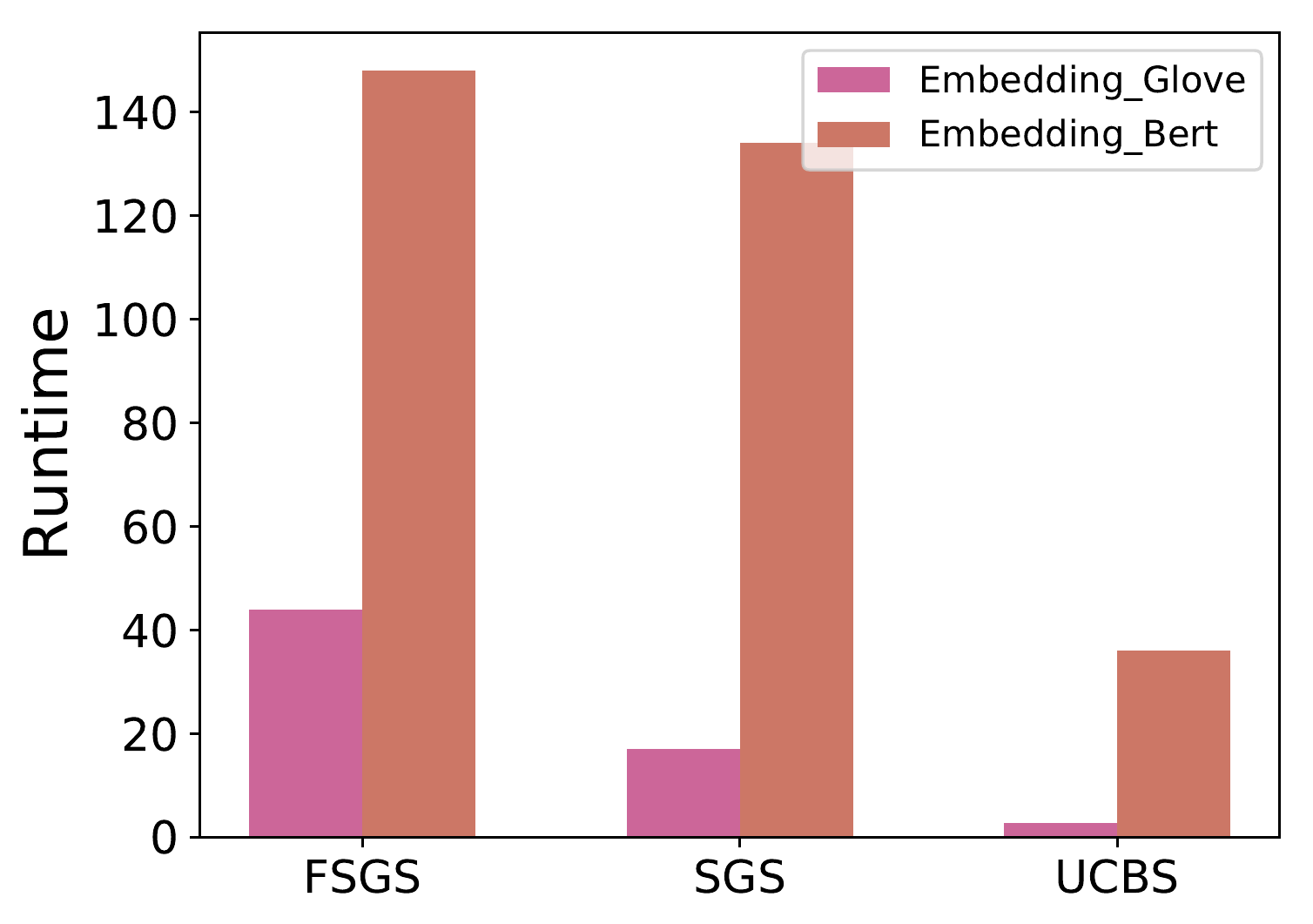}
\vspace{-2mm}
\caption{The assessment expenses on LIAR dataset of fakenews detection . Left: the number of queries. Right: average running time.}
\label{fig:LIAR_Runtime_app}
\vspace{-0.1cm}
\end{figure}

\textbf{\textit{Intrusion detection.}} 
In our study, \textit{AdvCat} conducts the robustness assessment task of \textit{Deeplog} on the HDFS and IPS dataset, following the setting in \cite{tiresias,deeplog,deepcase}. \textbf{{HDFS}} was collected by executing Hadoop-based map-reduce operations on over 200 Amazon EC2 nodes, and was labeled by intrusion detection domain experts. About 2.9 \% of the 11.2M system log entries gathered are anomalous\cite{deeplog}.
Our model was trained on 4,855 normal sessions, tested on 553,366 normal sessions and 15,200 abnormal sessions.
\textbf{{IPS}} \cite{tiresias} contains security events of a major security company's intrusion prevention product. Meta-information associated with a security event on an endpoint is recorded when this product detects network-level (e.g., unauthorized login) or system-level activity (e.g., malware detection) that matches a pre-defined signature. 
We use the anonymized machine ID to reconstruct a series of security events detected in a given machine and discard it after the reconstruction process is done.
We use the security events from 30k sessions recorded on a single day for training and 5k sessions of events for testing.
We strictly require that training and test data to come from different machines so as not to introduce evaluation bias~\cite{pendlebury2019tesseract}. Note that we measure both the top-1 and top-9 of the prediction as normal.
For all our experiments, we set the attack budget to 5, window size to 10, and the time limit to 60 seconds. More specifically, \textit{DeepLog} is performed over a sequence (a window) containing 10 logs. They use the 10 logs in this window and predict the successive log. 

\textbf{Modifying logs for robustness assessment.} \textit{AdvCat} applies modifications over the input logs to \textit{DeepLog}-based log prediction models for robustness assessment. Pioneering research efforts  \cite{Pierazzi2020IEEESP,Kathrin2017AdvMalware} in the security community unveil empirically modifying and/or removing code segments and network traffic packets can camouflage malicious incidents and evade ML-driven security event detection. In practices, an adversary has many hacking techniques, e.g. man-in-the-middle attacks and advanced persistent attacks to compromise the infrastructures transferring and/or storing these system event logs. He can then change/remove the logs to mislead the ML-based detection engine \cite{Pierazzi2020IEEESP}. 

\end{document}